\documentclass[confrence]{IEEEtran}
\IEEEoverridecommandlockouts
\usepackage{cite}
\usepackage{amsmath,amssymb,amsfonts}
\usepackage{algorithm}
\usepackage{algorithmic}
\usepackage{graphicx}
\usepackage{textcomp}
\usepackage{xcolor}
\usepackage{stfloats}
\usepackage{graphicx}
\usepackage{color}
\usepackage{multirow}
\usepackage{array}
\usepackage{bm}
\usepackage{subfig}
\usepackage{booktabs}

\newcommand{\tabincell}[2]{\begin{tabular}{@{}#1@{}}#2\end{tabular}}
\def\BibTeX{{\rm B\kern-.05em{\sc i\kern-.025em b}\kern-.08em
    T\kern-.1667em\lower.7ex\hbox{E}\kern-.125emX}}
\begin{document}
	
	\title{Joint Device Detection, Channel Estimation, and Data Decoding with Collision Resolution for MIMO Massive Unsourced Random Access\\}
	\author{\IEEEauthorblockN{Tianya Li, Yongpeng Wu, \emph{Senior Member, IEEE}, Mengfan Zheng, Wenjun Zhang, \emph{Fellow, IEEE}, \\Chengwen Xing, \emph{Member, IEEE}, Jianping An, \emph{Senior Member, IEEE}, Xiang-Gen Xia, \emph{Fellow, IEEE}, \\and Chengshan Xiao, \emph{Fellow, IEEE}}
		\thanks{T. Li, Y. Wu (corresponding author), and W. Zhang are with the Department of Electronic
			Engineering, Shanghai Jiao Tong University, Minhang 200240, China.
			Email: \{tianya, yongpeng.wu, zhangwenjun\}@sjtu.edu.cn.
			
			M. Zheng is with the Department of Electrical, and Electronic Engineering, Imperial College London, London SW7 2AZ, U.K. E-mail: m.zheng@imperial.ac.uk.
			
			C. Xing and J. An are with the School of Information and Electronics, Beijing Institute of Technology, Beijing 100081, China. E-mail: {xingchengwen@gmail.com}; an@bit.edu.cn. 
			
			X.-G. Xia is with the Department of Electrical, and Computer Engineering, University of Delaware, Newark, DE 19716 USA. E-mail: xianggen@udel.edu.
			
			C. Xiao is with the Department of Electrical, and Computer Engineering, Lehigh University, Bethlehem, PA 18015 USA. E-mail: xiaoc@lehigh.edu.
		}
	}

\maketitle
\begin{abstract}
\par In this paper, we investigate a joint device activity detection (DAD), channel estimation (CE), and data decoding (DD) algorithm for multiple-input multiple-output (MIMO) massive unsourced random access (URA). Different from the state-of-the-art slotted transmission scheme, the data in the proposed framework is split into only two parts. A portion of the data is coded by compressed sensing (CS) and the rest is low-density-parity-check (LDPC) coded. In addition to being part of the data, information bits in the CS phase also undertake the task of interleaving pattern design and channel estimation (CE). The principle of interleave-division multiple access (IDMA) is exploited to reduce the interference among devices in the LDPC phase. Based on the belief propagation (BP) algorithm, a low-complexity iterative message passing (MP) algorithm is utilized to decode the data embedded in these two phases separately. Moreover, combined with successive interference cancellation (SIC), the proposed joint DAD-CE-DD algorithm is performed to further improve performance by utilizing the belief of each other. Additionally, based on the energy detection (ED) and sliding window protocol (SWP), we develop a collision resolution protocol to handle the codeword collision, a common issue in the URA system. In addition to the complexity reduction, the proposed algorithm exhibits a substantial performance enhancement compared to the state-of-the-art in terms of efficiency and accuracy. 
\end{abstract}


\section{Introduction}

\subsection{Background and Related Works}
\par Next generation multiple access (NGMA) aims to support massive connectivity scenarios for many envisioned Internet of Things (IoT) applications, e.g. manufacturing, transportation, agriculture, and medicine, to be efficiently and flexibly connected \cite{wu2020wc,liu2018spm,Liu2018tsp,shyianov2021jsac,NGMA,B5G}. A generic scenario for IoT connectivity involves a massive number of machine-type connections \cite{liu2018spm}, known as the massive machine-type communication (mMTC), which is one of the main use cases of the fifth-generation (5G) and beyond \cite{B5G}. Compared to human-type communications (HTC), MTC has two distinct features: sporadic traffic and short packet communication. The sporadic access means that for any given time, for the sake of long battery life and energy saving, only a small fraction within a huge number of devices are active for data transmission \cite{Liu2018tsp}. While the small data payloads result in a fall of spectrum efficiency and an increase of latency when applied with traditional multiple access protocols. In this regard, a number of scalable random access schemes have been investigated in the content of massive connectivity. Conceptually, there are two lines of work for that end, namely, individual and common codebook-based approaches, respectively.

\par In the spirit of the individual codebook based approach, each device is equipped with a unique codebook for the identification conducted by the base station (BS) \cite{Liu2018tsp,gjy2021}. Since the BS is interested in the IDs of devices, this framework can be referred to as sourced random access (SRA). A typical transmission process includes two phases in SRA, namely, training and data transmission, respectively. In the training phase, the BS conducts the task of device activity detection (DAD) and channel estimation (CE) based on the pre-allocated unique pilot sequences transmitted by active devices, while the data transmission is performed in the following phase. The task of DAD-CE can be modeled as the compressed sensing (CS) problem and there have been quantities of algorithmic solutions for this issue. Among these CS recovery techniques, the approximate message passing (AMP) algorithm, which is first proposed in \cite{donoho2009message} for the single measurement vector (SMV) based problem, exhibits great performance in terms of accuracy and complexity. Since the publication of \cite{donoho2009message}, there have been vast variants based on the AMP algorithm, such as multiple measurement vector (MMV) AMP \cite{Liu2018tsp}, generalized MMV (GMMV) AMP \cite{ke2020tps}, orthogonal AMP (OAMP) \cite{ma2017oamp}, generalized AMP (GAMP) \cite{GAMP} and many other works. Another line of work is based on the belief propagation (BP) algorithm \cite{BP}, which models the problem as a graph and iteratively calculates the messages among different nodes \cite{zhang2020iot,hiroki2020arxiv,yuan2020twc,huang2019tsp}. Thanks to the consistency of the message updating rules, the messages can always be jointly updated within iterations to obtain improved performance. However, the premise of these algorithms is that each device should have a unique pilot sequence. As we jump out of these algorithmic solutions and go back to the essence of the individual codebook, we find this is inapplicable to assign each device a unique pilot in the mMTC scenario with a huge amount of potential devices. Meanwhile, it is a waste of resources especially with sporadic traffic. To mitigate this issue, the study on the framework of common codebooks is rapidly developed.

\par As opposed to the case of individual codebooks, active devices choose codewords from a common codebook for their messages. In this regard, the task of BS is only to produce the transmitted messages regardless of the corresponding IDs, thus leading to the so-called unsourced random access (URA). The main difference between URA with other grant-based and grant-free random access protocols is that the BS does not perform device identification but only decodes the list of active device messages up to permutations \cite{B5G}. Conceptually, there are two distinct advantages for this novel framework: $\left.i\right)$ the BS is capable of accommodating massive devices, since all devices share the common codebook instead of being assigned unique preamble sequences as in traditional schemes. $\left.ii\right)$ No device identification information is needed to be embedded in the transmitted sequence. As such, the overhead will be reduced, contributing to improved efficiency. The study on URA is first reported in \cite{yury2017isit}, in which Polyanskiy considered the scenario of a massive number of infrequent and uncoordinated devices and discussed a random coding existence bound in the Gaussian multiple-access channel (GMAC). Since the publication of \cite{yury2017isit}, there have been many works devoted to approaching that bound \cite{or2017isit,Amalladinne2020tit,Fengler2019isit,sparcs2021tit,polar2020icc,polar2020vtc,Vem2019gc,Vem2019tcom}. A low complexity coding scheme is investigated in \cite{or2017isit}. Based on the combination of compute-and-forward and coding for a binary adder channel, it exhibits an improved performance compared with the traditional schemes, such as slotted-ALOHA and treating interference as noise (TIN). However, the size of the codebook increases exponentially with the blocklength, resulting in difficulty in achieving that underlying bound. To mitigate this issue, a slotted transmission framework is proposed in \cite{Amalladinne2020tit}, referred to as the coded compressed sensing (CCS) scheme. This cascaded system includes inner and outer codes. The outer tree encoder first splits the data into several slots with parity bits added. The CS coding is then employed within each slot to pick codewords from the codebook. The blocklength, as well as the size of the codebook in each slot, are greatly reduced, thus leading to a relaxation of the computational burden. On the receiver side, the inner CS-based decoder is first utilized to recover the slot-wise transmitted messages within each slot. The outer tree decoder then stitches the decoded messages across different slots according to the prearranged parity. This structure is inherited by the later works in \cite{Fengler2019isit,sparcs2021tit}, where the authors exploit the concept of sparse regression code (SPARC) \cite{sparcs2012} to conduct the inner encoding and decode it with the AMP algorithm. Some other coding schemes for URA are also reported in \cite{polar2020icc,polar2020vtc,Vem2019gc,Vem2019tcom}. The polar coding schemes based on TIN and successive interference cancellation (SIC) are investigated in \cite{polar2020icc,polar2020vtc}. In \cite{Vem2019gc,Vem2019tcom}, the author consider a hierarchical framework, where the device's data is split into two parts. A portion of the data is encoded by a common CS codebook and recovered by the CS techniques, while the rest is encoded by the low-density parity check code (LDPC) \cite{LDPC} with the key parameters conveyed in the former part. Besides the above works considering the GMAC, there are also works in the fading channel \cite{fading2019acssc,fading2019isit,fading2020tcom}.

\par Besides the above works considering a single receive antenna at the BS, the study of massive URA in multiple-input multiple-output (MIMO) systems has also drawn increasing attention. The BS equipped with multiple antennas provides extra dimensions for the received signal, thus offering more opportunity for signal detection and access of a massive number of devices. A coordinated-wise descend activity detection (AD) algorithm is proposed in \cite{Haghighatshoar2018isit,Fengler2021tit}, which finds the large-scale fading coefficients (LSFCs) of all the devices (coordinates) iteratively. The authors adopt a concatenated coding scheme with the aforementioned tree code as the outer code, and a non-Bayesian method (i.e., maximum likelihood (ML) detection) is leveraged as the inner decoder based on the covariance of the received signal, referred to as the covariance-based ML (CB-ML) algorithm. However, the performance of the CB-ML algorithm degrades dramatically when the number of antennas is less than that of the active devices. There are also some improvements to this algorithm in terms of complexity or accuracy \cite{Shyianov2021jsac,fast}. The slotted transmission framework proposed in \cite{Shyianov2021jsac} eliminates the need for concatenated coding. That is, no parity bit is added within each slot and the slot-wise messages can be stitched together by clustering the estimated channels. However, the channels in \cite{Shyianov2021jsac} are assumed to be identically distributed over all the slots, which is difficult to hold in practice. Besides, the collision in codewords (more than one device chooses the same codeword) will lead to a superimposition of the estimated channels, resulting in a failure of the stitching process. Currently, the design of an efficient collision resolution scheme in URA remains missing. In \cite{fast}, a more efficient coordinate selection policy is developed based on the multi-armed bandit approaches, leading to a faster convergence than the CB-ML algorithm. Nevertheless, these slotted transmission frameworks, such as CB-ML and its variants, all map a piece of data within each slot to a long transmitted codeword based on the CS coding, resulting in low efficiency. In this regard, the algorithmic design for the MIMO massive URA with high efficiency and accuracy remains missing.
 
\subsection{Challenges}
\par In this paper, we investigate a two-phase transmission scheme for MIMO massive URA, where the device's transmitted data is split into two phases with CS and LDPC coded, respectively. In this framework, no parity bits are needed to be embedded in the CS phase, and the data is encoded linearly in the LDPC phase instead of being mapped to a long codeword. As such, it exhibits higher efficiency and lower latency than the aforementioned slotted transmission scheme. The existing schemes considering this framework all focus on the single-antenna case. For instance, Polar coding \cite{polar2020vtc} and LDPC coding schemes \cite{Vem2019gc,Vem2019tcom} consider the GMAC with perfect channel state information (CSI) at the BS. In this regard, the above channel coding schemes can obtain satisfying performance. However, such an assumption cannot be established under the MIMO case. Besides, because of the massive connectivity and sporadic traffic of devices, the positions of active devices in the estimated equations are not determined. Consequently, conventional linear receivers, such as zero-forcing (ZF), are not applicable. Moreover, compared with the fading channel models \cite{fading2019acssc,fading2019isit,fading2020tcom} where the channel of each device is a scalar, in MIMO, the channel is a vector, and elements among antennas share the same activity. Therefore, elements in the channel vector should be estimated jointly rather than separately, of which the distribution is formulated as a multi-dimensional Bernoulli-Gaussian distribution in this paper. In this regard, the conventional estimators, such as the minimum mean square error (MMSE) estimation or the maximum \emph{a posteriori} (MAP) estimation, are hard to carry out straightforwardly. Specifically, it is computationally intractable to obtain a precise posterior distribution, since it involves marginalizing a joint distribution of the activity and channel with high dimensions.

\par Another challenge for the proposed scheme in the MIMO case is activity detection. We note that whether in the GMAC \cite{polar2020vtc,Vem2019gc,Vem2019tcom} or fading channel \cite{fading2019acssc,fading2019isit,fading2020tcom} in the single-antenna case, the device activity can be directly detected. For instance, since the device's channel is a scalar in the fading channel, it can be simply determined to be active when the estimated channel is non-zero or larger than a given threshold. However, it will be much involved in MIMO because there are multiple observations of the activity in the estimated channel vector. The existing works, such as \cite{ke2020tps}, make a hard decision based on the energy of the channel and \cite{Shyianov2021jsac} considers the LLR at each antenna separately and sums them to obtain the final decision. However, the threshold for the decision is hard to obtain in practice and the distribution of the channel is not utilized in \cite{ke2020tps}. Although \cite{Shyianov2021jsac} considers the channel distribution and gives a closed-form expression for activity detection, the nature that channels at each antenna share the same activity is not considered.

\par Besides, the LDPC code is efficient but sensitive to the CE results. Therefore, the overall performance is limited by the accuracy of CE. Moreover, with the presence of collision in URA, if more than one device chooses the same codeword in the CS phase, the corresponding channels will be superimposed and thus the subsequent LDPC decoding process will fail.

\subsection{Contributions}
\par To cope with the arising issues, based on the message passing (MP) algorithm, we propose the Joint DAD-CE algorithm to conduct the task of joint activity detection and channel estimate in the CS phase, and MIMO-LDPC-SIC Decoding algorithm for data decoding embedded in the LDPC phase. Moreover, to further improve the performance, we propose the Joint DAD-CE-DD algorithm to jointly update the messages in these two phases by utilizing the belief of each other. Finally, we propose a collision resolution protocol to address the collision in URA. The key and distinguishing features of the proposed algorithms are listed below.

\begin{itemize}
	\item Based on the principle of the BP algorithm, we develop a low-complexity iterative MP algorithm to decode the two parts of data. For the CS phase, we investigate the Joint DAD-CE algorithm to recover part of the data embedded in the device activities, the interleaving patterns, and channel coefficients, which are the key parameters for the remaining data. Specifically, we derive a close-form expression for DAD by utilizing the joint distribution of the channel among antennas. Combined with SIC and in the spirit of interleave-division multiple access (IDMA) \cite{ping2006twc}, we elaborate on the MIMO-LDPC-SIC Decoding algorithm to decode the remaining data embedded in the LDPC phase. In addition to complexity reduction, the proposed algorithm exhibits a substantial performance enhancement in terms of accuracy and efficiency compared to the state-of-the-art CB-ML algorithm.
	
	\item Thanks to the consistency of the MP algorithm, we propose the Joint DAD-CE-DD algorithm to further improve the performance. The proposed algorithm suggests a paradigm connecting the two parts. That is, messages in the decoding process of CS and LDPC parts can be jointly updated by utilizing the belief of each other, thus leading to improved performance. We employ the correctly decoded codewords as soft pilots to conduct CE jointly with the codewords in the CS phase, which contributes to improved accuracy of the estimated channel. Combined with the SIC method, the accuracy of the residual signal can be improved, leading to the enhanced decoding performance.
	
	\item Under the current framework based on the common codebook, a collision happens if there are more than one device having the same preamble, which leads to a superimposition of the estimated channels. Accordingly, the superimposed channel will cause a failure in the LDPC decoding process. To this end, we propose a collision resolution protocol based on the energy detection (ED) and sliding window protocol (SWP). Succinctly, the ED is performed on the estimated channel of each device to find out the superimposition. Then, the BS broadcasts the indexes of the superimposed channels to all devices. Afterwards, the devices in collision slide the window in the data sequence and the CS coding is again performed for retransmission.
\end{itemize}

\par The rest of the paper is organized as follows. In Section \ref{sec-2}, we introduce the system model for MIMO massive URA . In Section \ref{sec-3}, we implement the two-phase encoding scheme and the collision resolution protocol is developed in Section \ref{sec-4}. Then, we elaborate on the low-complexity iterative decoding algorithm based on BP and explain the joint update algorithm in Section \ref{sec-5}. After verifying the numerical results and analyzing the complexity in \ref{simulation}, we conclude the paper in Section \ref{sec-7}.

\subsection{Notations}
\par Unless otherwise noted, lower- and upper-case bold fonts, $\bm{x}$ and $\bm{X}$, are used to denote
vectors and matrices, respectively; the $(m,n)$-th entry of $\bm{X}$ is denoted as ${X}_{m,n}$; $\bm{X}_{i,:}$ denotes the $i$-th row of $\bm{X}$; $\{\cdot\}^*$ denotes the conjugate of a number; $\{\cdot\}^T$ and $\{\cdot\}^H$ denote transpose and conjugate transpose of a matrix, respectively; $\mathbb{E}\{\cdot\}$ and $\text{Var}\left\lbrace\cdot \right\rbrace $ denote the statistical expectation and variance, respectively; $\bm{X}\sim\mathcal{CN}(\bm{x};\bm{\mu},\bm{\Sigma})$ means that the random vector $\bm{x}$ follows a complex Gaussian distribution with mean $\bm{\mu}$ and auto-covariance matrix $\bm{\Sigma}$; $\{\cdot\}!$ denotes the factorial; $\left[1:M \right] $ denotes the set of integers between $1$ and $M$; $ \mathcal{L} \backslash l$ denotes the entries in set $\left\lbrace 1,2,\cdots,L\right\rbrace $ except $l$; $\text{Re}(\cdot)$ and $\text{Im}(\cdot)$  denote the real and imaginary parts of a complex number, respectively; $\lceil \cdot  \rceil$ denotes ceiling; $i$ is the imaginary unit (i.e., $i^2=-1$).
 
\section{System Model} \label{sec-2}

\par Consider the uplink of a single-cell cellular network consisting of $K_{tot}$ single-antenna devices, which are being served by a base station (BS) equipped with $M$ antennas. This paper assumes sporadic device activity, i.e, a small number, $K_a \ll K_{tot}$ of devices are active within a coherence time. Each device has $B$ bits of information to be coded and transmitted into a block-fading channel with $L$ channel uses. Let $\bm{v}_k \in \left\lbrace 0,1\right\rbrace ^B$ denote device $k$'s binary message and $f(\cdot): \left\lbrace 0,1\right\rbrace ^B \rightarrow \mathbb{C}^L$ is some one-to-one encoding function. Typically, in URA scenario, the implementation of $f(\cdot)$ is to select the corresponding codeword $\bm{a}_k$ from a shared codebook $\bm{A}=\left[ \bm{a}_1,\bm{a}_2,\cdots,\bm{a}_{2^{B}} \right] \in \mathbb{C}^{L \times2^B}$ according to $\bm{v}_k$ \cite{Haghighatshoar2018isit,Fengler2019isit,Fengler2021tit}. The corresponding received signal can be written as 
\begin{equation}
	\bm{Y}= \sum\nolimits_{k \in \mathcal{K}_{tot}}{\phi_k f(\bm{v}_k) \tilde{\bm{h}}_k^T} +\bm{Z}, \label{equ-1}
\end{equation}
where $\phi_k$ is the device activity indicator, which is modeled as a Bernoulli random variable and defined as follows:
\begin{equation}
	\phi_{k}=\left\{\begin{array}{l}
		1, \text { if device } k \text { is active, } \\
		0, \text { otherwise }
	\end{array} \quad \forall k \in \mathcal{K}_{tot}.\right. \label{equ-2}
\end{equation}
 $\tilde{\bm{h}}_k \in \mathbb{C}^{M \times 1}$ is the channel vector of device $k$ and $\bm{Z}\in \mathbb{C}^{L\times M}$ is the additive white Gaussian noise (AWGN) matrix distributed as $\mathcal{CN}(0,\sigma^2\bm{I}_M)$. In line with the state-of-the-art setting \cite{Fengler2021tit}, we assume for simplicity and fair comparison that $\bm{h}_k$ are i.i.d., i.e., independent of each other and uncorrelated among antennas. Specifically, the Rayleigh fading model is considered in this paper, where $\tilde{\bm{h}}_k=\sqrt{\beta_{k}}\bm{g}_{k}$ and $\bm{g_{k}}\sim \mathcal{CN}(0,\bm{I}_M)$ denotes the Rayleigh fading component, and $\beta_{k}$ denotes LSFC. We would like to mention that more realistic channel models in MIMO or massive MIMO have been discussed and well investigated, such as the spatially correlated fading channel \cite{ke2020tps,add-corre,gao2015tsp,you2015twc,xxy2021iccc}. Due to the limited angular spread, the channel in the virtual angular domain exhibits a sparsity among antennas. As such, the data of devices is not directly superimposed but staggered on different antennas, and the multi-user interference can be further reduced, contributing to improved performance. Nevertheless, for the sake of consistency with the benchmarks \cite{Liu2018tsp,Fengler2021tit} and isolating the fundamental aspects of the problem without additional model complication, we embark throughout this paper on the study of i.i.d. Rayleigh fading channel. The study on the spatially correlated channel is remarkable and left for future work.

\par The BS's task is to produce a list of transmitted messages  $\mathcal{L}(\bm{Y})$  without identifying from whom they are sent, thereby leading to the so-called URA. The performance of a URA system is evaluated by the probability of missed detection and false alarm, denoted by $p_{md}$ and $p_{fa}$, respectively, which are given by:
\begin{align}  
	\label{equ-3} p_{md} &= \frac{1}{{{K_a}}}\sum\nolimits_{k \in {\mathcal{K}_a}} {P\left( {{\bm{v}_{{k}}} \notin \mathcal{L}} \right)} \\
	\label{equ-4} p_{fa} &= \frac{{\left|  {\mathcal{L}\backslash \left\{ {{\bm{v}_{{k}}}:k \in {\mathcal{K}_a}} \right\}} \right|}}{{\left| \mathcal{L} \right|}}.
\end{align} 
In this system, the code rate $R_c={B}\slash{L}$ and the spectral efficiency $\mu=\frac{B\cdot K_a}{L \cdot M}$. Let $P$ denote the power (per symbol) of each device, then the energy-per-bit $E_b/N_0$ is defined by
\begin{equation}
	E_b/N_0 = \frac{LP}{2B}.
\end{equation}

 \section{Encoder} \label{sec-3}
 \par
\begin{figure}[htpb]
 	\centerline{\includegraphics[width=0.4\textwidth]{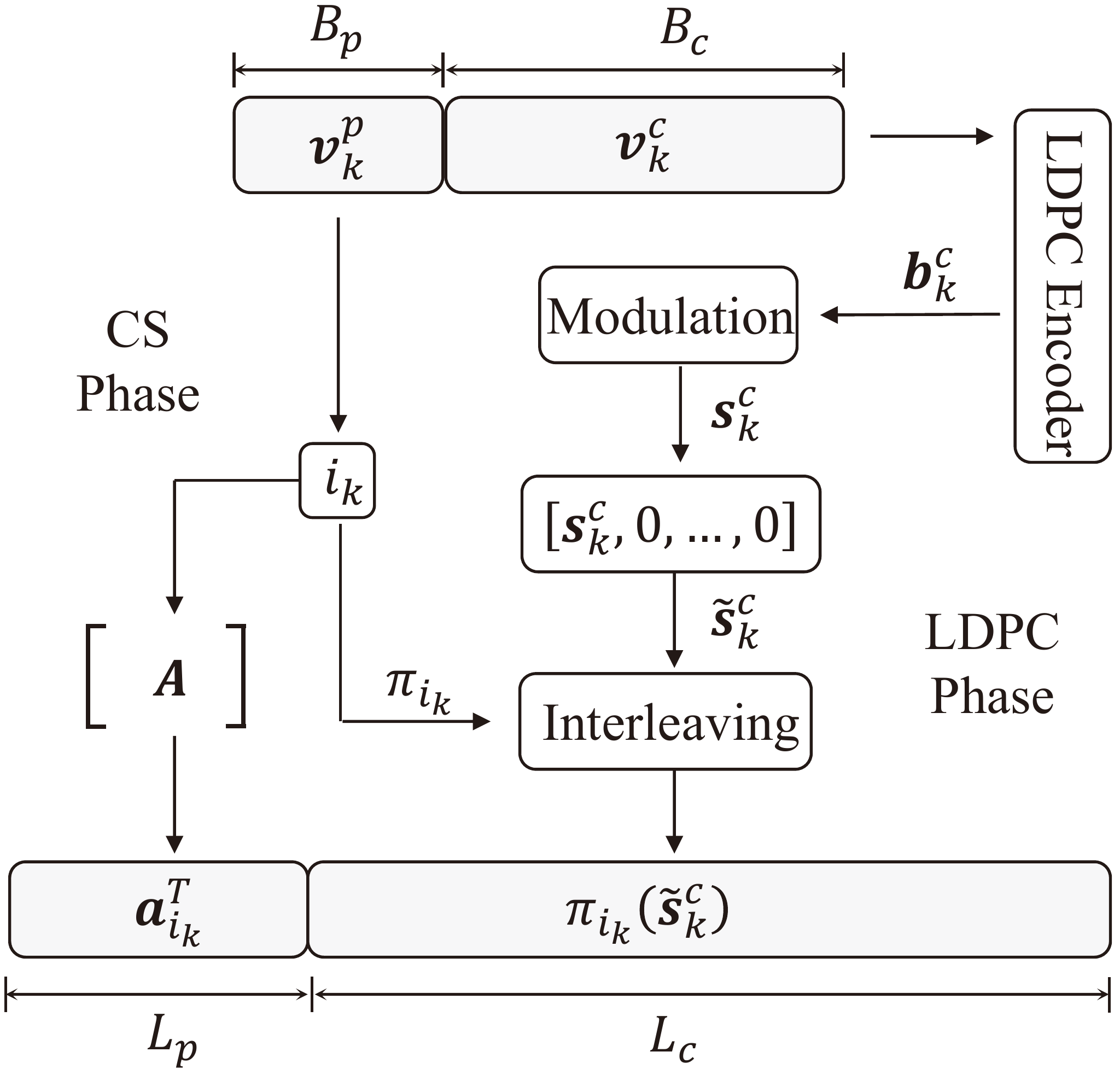}}
 	\caption{The encoding process in the CS and LDPC phases.}
 	\label{pic-1}
 \end{figure}

 As aforementioned, the existing slotted transmission scheme exhibits low efficiency by the CS coding. To mitigate this issue, a two-phases coding scheme is proposed in \cite{Vem2019tcom,Vem2019gc}, which considers the $T$-user Gaussian multiple access (GMAC) channel and also URA scenario. We extend this work to the MIMO system and refer to it as the CS-LDPC coding scheme. Similarly, the hierarchical form of the encoding process is considered in this paper. The $B$ bits of information are first divided into two parts, namely, $B_p$ and $B_c$ bits with $B_p+B_c=B$. Typically, one would want $B_p \ll B_c$. The former $B_p$ information bits are coded by a CS-based encoder to pick codeword from the common codebook. Based on the codebook, the BS is tasked to recover part of the messages, the number of active devices, channel coefficients as well as interleaving patterns for the latter part. The remaining $B_c$ bits are coded with LDPC codes. For clarity, we denote the former and latter encoding processes as CS and LDPC phases, respectively. Correspondingly, the total $L$ channel uses are split into two segments of lengths $L_p$ and $L_c$, respectively, with $L_p+L_c=L$. Since only a small fraction of data is CS coded and the rest is LDPC coded, the efficiency in our scheme is higher than those purely CS-coded schemes \cite{Shyianov2021jsac,Amalladinne2020tit,Fengler2019isit,Fengler2021tit}. The key features of this encoding process are summarized in Fig. \ref{pic-1}. We elaborate on these two encoding phases below.
 
 \subsection{CS Phase} \label{sec3-1}
 \par The URA fashion is considered in this phase. Let $\mathbf{A}\in \mathbb{C}^{L_p \times 2^{B_p}}$ denote the common codebook shared by all the devices. That is, the columns of $\mathbf{A}=\left[ \bm{a}_1,\bm{a}_2,\cdots,\bm{a}_{2^{B_p}} \right] $ form a set of codewords with power constraint $\left\| {{\mathbf{a}_{i}}} \right\|_2^2 = L_p$, from which each active device chooses in order to encode its ${B_p}$ bits of information. With a slight abuse of notation, let $\bm{v}_k^p$ denote the first $B_p$ bits of device $k$'s binary message, i.e., $\bm{v}_k^p \triangleq \bm{v}_k(1:B_p) \in \left\lbrace 0,1\right\rbrace ^{B_p}$. To apply the encoding scheme, we convert $\bm{v}_k^p$ into the $\ell_1$-norm  binary vector $\tilde{\bm{v}}_k^p \in \left\lbrace 0,1\right\rbrace ^{2^{B_p}}$, in which a single one is placed at the location $i_k$. The value $i_k$ of binary sequence $\bm{v}_k^p$ is obtained by regarding it as an integer expressed in radix-$2$ (plus one), which we write as $i_k=\left[ \bm{v}_k^p \right]_2 \in \left[ 1: 2^{B_p} \right] $. Then, the coded sequence is obtained by taking the transpose of the $i_k$-th column of $\bm{A}$, which we denote by $ \bm{a}_{i_k}^T$. This facilitates the CS architecture, which maps the information to an elementary vector $\tilde{\bm{v}}_k^p$, according to which the corresponding codeword is selected from a fixed codebook. The corresponding received signal can be written as
 \begin{equation}
 	\bm{Y}= \sum\nolimits_{k \in \mathcal{K}_{tot}}{\phi_k \bm{a}_{i_k}\tilde{\bm{h}}_k^T} +\bm{Z}, \label{equ-5}
 \end{equation}
 where $\phi_k$ is the device activity indicator, as defined in (\ref{equ-2}). The matrix form of (\ref{equ-5}) is given by
 \begin{equation}
 	\bm{Y}=\bm{A}\bm{\Gamma}\tilde{\bm{H}}+\bm{Z}, \label{equ-6}
 \end{equation}
 where $\mathbf{A}\in \mathbb{C}^{L \times 2^{B_p}}$ is the common codebook shared by all devices. $\mathbf{\Gamma}  \in {\left\{ {0,1} \right\}^{{2^{B_p}} \times {K_{tot}}}}$ denotes the binary selection matrix. For each active device $k\in \mathcal{K}_{a}$, the corresponding column $\bm{\Gamma}_{:,k}$ is all-zero but a single one in position $i_k$, while for all $\mathcal{K}_{tot}  \backslash \mathcal{K}_a$ the corresponding column $\bm{\Gamma}_{:,k}$ is all zeros. $\tilde{\bm{H}}=\left[ \tilde{\bm{h}}_1,\cdots,\tilde{\bm{h}}_{K_{tot}}\right]^T \in \mathbb{C}^{K_{tot}\times M}$ corresponds to the MIMO channel coefficient matrix. We note that the number of total devices $K_{tot}$ plays no role in URA. In order to get rid of this variable, with slight abuse of notation we define the modified activity indicator and selection matrix as $\phi_{r}=\sum\nolimits_{k\in \mathcal{K}_a}{\gamma_{r,k}}$ and $\bm{\Phi}=\text{diag}\left\lbrace \phi_1, \cdots, \phi_{2^{B_p}}\right\rbrace$, respectively, where $\gamma_{r,k}$ is the $(r,k)$-th entry of $\bm{\Gamma}$. Correspondingly, the modified channel is ${\bm{H}}=\left[\bm{h}_1,\cdots, \bm{h}_{2^{B_p}} \right]$ with $\bm{h}_r=\sum\nolimits_{k \in {\mathcal{K}_a}}{\gamma_{r,k}\tilde{\bm{h}}_k}$. Hence, \eqref{equ-6} can be written as 
 \begin{equation}
 	\bm{Y}=\bm{A}\bm{\Phi}\bm{H}+\bm{Z}. \label{equ-6-add}
 \end{equation}
  Let $\bm{X}=\bm{\Phi}\bm{H}=\left[ \bm{x}_1,\cdots,\bm{x}_{2^{B_p}}\right]^T $. The goal for the BS in the CS phase is to detect the non-zero positions of the selection matrix $\bm{\Phi}$ and the corresponding channel vectors by recovering $\bm{X}$ based on the noisy observation $\bm{Y}$. Since $\bm{X}$ is row sparse, i.e., many $\bm{x}_n$ are zero, such reconstruction problem can be modeled as the CS problem \cite{Liu2018tsp}. Once $\bm{\Phi}$ is recovered, the message indicators of active devices $\left\lbrace i_k, \forall k\in \mathcal{K}_a \right\rbrace $ are also recovered. Moreover, $i_k$ acts as the parameter of the LDPC code, since it determines the interleaving pattern of the data in the LDPC phase. Considering the fact that the $B_p$ bits of message carries key parameters for the latter phase, we name it the preamble. We note that it is different from the preamble in the traditional grant-free scenario, which is a pure pilot with no data embedded.

 \subsection{LDPC Phase}
 \par Likewise, let $\bm{v}_k^c \triangleq \bm{v}_k(B_p+1:B) \in \left\lbrace 0,1\right\rbrace ^{B_c}$ denote the remaining $B_c$ information bits of device $k$. $\bm{v}_k^c$ is first encoded into an LDPC code $\bm{b}_k^c \in \left\lbrace 0,1 \right\rbrace^{\tilde{L}_c}$, which is determined by the LDPC parity check matrix $\bm{C}$ with size $\left(\tilde{L}_c-B_c \right)\times \tilde{L}_c$. We note that in the decoding process, if $\hat{\bm{v}_k^c}$ is a valid codeword, then  $\text{mod} \left( \bm{C}\hat{\bm{v}}_k^c,2\right) =0$. $\bm{b}_k^c$ is then modulated to $\bm{s}_k^c$. We adopt a sparse spreading scheme introduced in \cite{Vem2019gc}. That is, $\bm{s}_k^c$ is zero-padded into a length $L_c$ vector $\tilde{\bm{s}}_k^c$
 \begin{equation}
 	\tilde{\bm{s}}_k^c = \left[\bm{s}_k^c, ~0, ~\cdots, ~0 \right]. \label{equ-7}
 \end{equation}
 We then employ the index representation $i_k$ to permute the ordering of $\tilde{\bm{s}}_k^c$. This is implemented by a random interleaver $\pi_{i_k}$ with interleaving pattern $i_k$. As mentioned in \cite{Vem2019tcom}, the purpose for permuting the codewords is to decorrelate the random access interference from other devices. This is similar to the IDMA scheme since the interleaving patterns for most of the devices are different because of the distinctive indices $i_k$. Hence, $\bm{v}_k^c$ is finally encoded to $\pi_{i_k}\left( \tilde{\bm{s}}_k^c\right)$. Appending it to the coded message in the CS phase yields the final codeword $\bm{x}_k$:
 \begin{equation}
 	\bm{x}_k = \left[ \bm{a}_{i_k}^T, \pi_{i_k}\left( \tilde{\bm{s}}_k^c\right) \right]^T. \label{equ-8}
 \end{equation}

\par The received signal including both the CS and LDPC phases is given by
 \begin{equation}
 	\bm{Y}=\sum\nolimits_{k \in \mathcal{K}_{a}}{\bm{x}_k \bm{h}_{i_k}^T} + \bm{Z}, \label{equ-9}
 \end{equation}
where $\bm{h}_{i_k}$ is assumed to follow independent quasi-static flat fading within the above two phases in this paper. The goal for the BS is to recover $\left\lbrace  \bm{v}_k, \forall k\in \mathcal{K}_a \right\rbrace $ based on the received signal $\bm{Y}$ and the channel $\bm{h}_{i_k}$  estimated in the CS phase. We emphasize again that the BS only produces the transmitted messages without distinguishing the corresponding devices. 

\par This hierarchical encoding process appears to be similar to the work of \cite{Vem2019tcom}, where CS and channel coding techniques are utilized to encode the messages. However, unlike in \cite{Vem2019tcom} considering the GMAC system, our approach is in the MIMO channel. Hence, in addition to recovering the parameters of LDPC codes conveyed by the codebook $\bm{A}$ as in \cite{Vem2019tcom}, the BS is also tasked to estimate the channel coefficients. Besides, as we will see shortly in Section \ref{sec-5-3}, a belief propagation decoder draws a connection between the CS and LDPC phases. That is, messages in the CE as well as DD processes can be jointly updated by utilizing the belief of each other. Whereas the two phases in \cite{Vem2019tcom} are two independent modules and work sequentially.
 
 \section{Collision Resolution Protocol} \label{sec-4}
  
 \par It is possible that two or more devices have the same preamble message, $\bm{v}^p$, which, although, may occur in a  small probability. In this case, the collided devices will have the same interleaving pattern in the LDPC phase, which goes against the principles of the IDMA scheme. Moreover, they will choose the same codeword as their coded messages in the CS phase, which results in the received signal in the CS phase being
 \begin{equation}
 	\bm{Y}_{colli} = \sum\nolimits_{i_k\in\mathcal{C}}{\bm{a}_{i_k}\sum\nolimits_{j \in \mathcal{K}_{i_k}}{\tilde{\bm{h}}_j^T}} + \bm{Z}, \label{equ-10}  
 \end{equation}
where $\mathcal{C}$ and $\mathcal{K}_{i_k}$ denote the set of collided indexes and collided devices corresponding to the message index $i_k$, respectively. In this regard, the BS can only recover the superimposed channels of these devices, instead of their own, which will lead to the failure of the LDPC decoding process. Although collision may occur in a small probability, it can occur. However, the design of an efficient collision resolution scheme in URA remains missing.

\begin{figure*}[t]
	\centering
	\subfloat[The BS broadcasts collided index representations $\left\lbrace i_k | \epsilon_{i_k}>\eta  \right\rbrace $ to all the devices. Based on this, active devices can figure out whether they have been in a collision. Deivce $i$ to $j$ are in collision in this case.]{\includegraphics[width=0.38\linewidth]{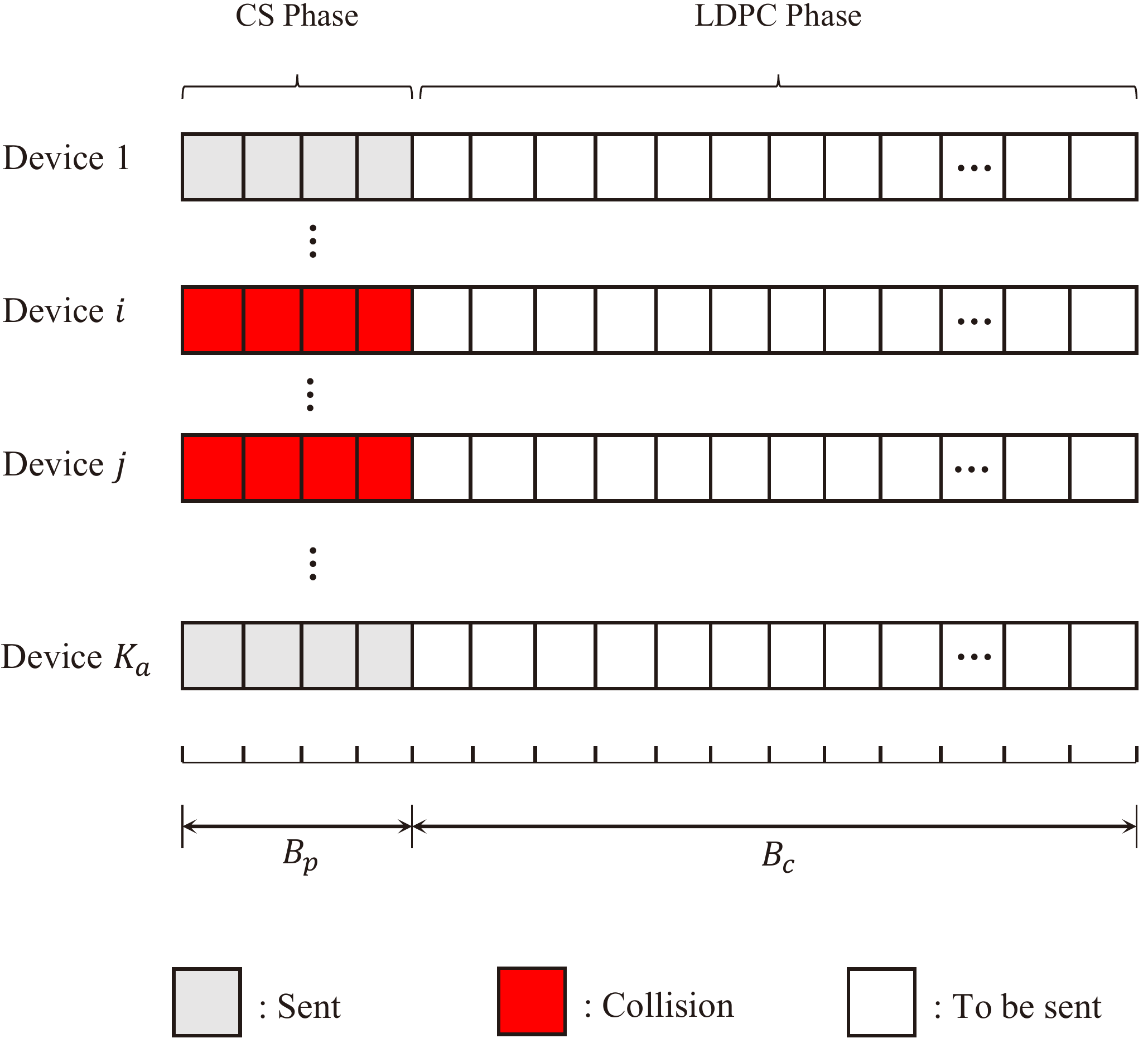} \label{pic2-1}}
	\hfil
	\subfloat[The collided devices slide the window with length $B_p$ bits forward within the total $B$ bits to get new sequences. The sliding length is $B_0$, satisfying $0<B_0<B_p$.]{\includegraphics[width=0.38\linewidth]{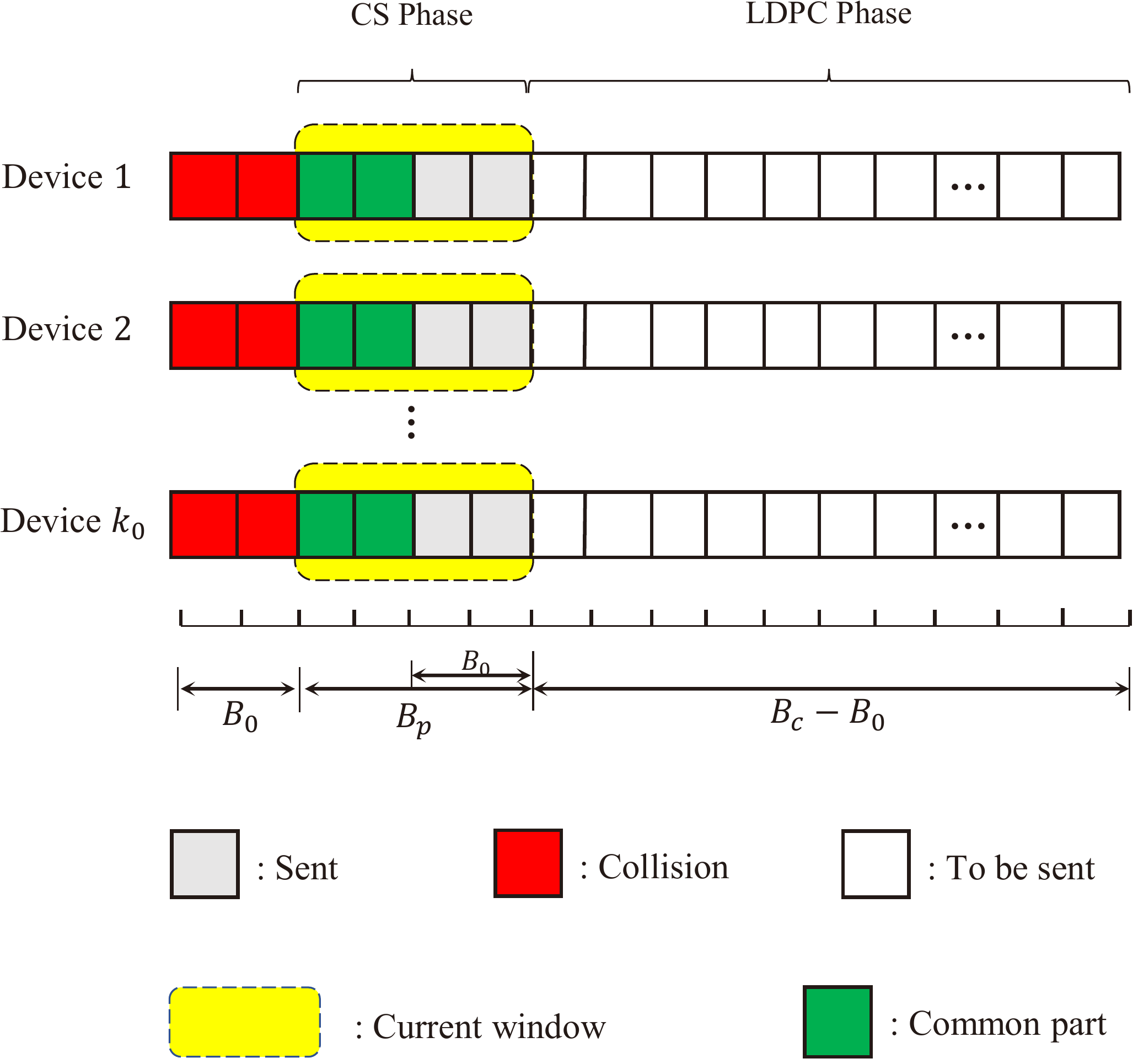}\label{pic2-2}}
	\caption{The diagram of the collision resolution protocol.}
\end{figure*}

\par In this paper, we develop a collision resolution protocol based on the ED and SWP. We note that in real scenarios, the near-far effects can be well solved with the existing power control schemes \cite{ Chandrasekhar2009twc,Patel1994jsac,Turin1984jsac}. In this regard, a flat fading channel is considered in this paper. That is, the LSFCs $\beta_k$ for all the devices are assumed to be identical, as also considered in \cite{Haghighatshoar2018isit,Fengler2021tit}. As mentioned above, if a collision happens, the recovered channels of the collided devices will be a superposition of their own, that is 
\begin{equation}
	\hat{\bm{h}}_{i_k} = \sum\nolimits_{j \in \mathcal{K}_{i_k}}{\tilde{\bm{h}}_j} + \bm{z}, \quad i_k\in \mathcal{C}, \label{equ-11}
\end{equation}
where $\bm{z} \sim \mathcal{CN}(0,\sigma^2\bm{I}_M)$ denotes the Gaussian noise. And $\hat{\bm{h}}_{i_k}$ is distributed as $\hat{\bm{h}}_{i_k} \sim \mathcal{CN}(0,\left( |\mathcal{K}_{i_k}|+\sigma^2)\bm{I}_M\right) $, which has a higher power than those without collision. Therefore, an effective way to detect collision is to perform ED on the estimated channel by the BS
\begin{equation}
	\epsilon_{i_k} = \mathbb{E}\left[ \hat{\bm{h}}_{i_k}^H \hat{\bm{h}}_{i_k} \right]. \label{equ-12}
\end{equation}
If $\epsilon_{i_k}$ is greater than a given threshold $\eta$, then it is utilized as evidence that there are at least two devices that have the same preamble and thus they choose the same CS codeword $\bm{a}_{i_k}$. Since devices themselves do not know whether they have been in a collision, the BS needs to feed this information back. To this end, the BS will broadcast the collided index representations $\left\lbrace i_k | \epsilon_{i_k}>\eta  \right\rbrace $ to all the devices to help them get the judgment. Fig. \ref{pic2-1} showcases that device $i$ to $j$ realized that a conflict occurred after receiving the indexes broadcast by the BS. We note that in the above process, the additional information required at the BS is only the threshold and it can be easily preset in practice.

\par As illustrated in Fig. \ref{pic2-2}, in order to get a new non-conflicting index representation, the collided devices will slide the window with length $B_p$ bits forward within the total $B$ bits to get new sequences, denoted by ${\bm{v}_k^p}'$. We denote $B_0$ as the sliding length which satisfies $0<B_0<B_p$. The reason behind $B_0<B_p$ is that there should be a common part, labeled in green in Fig. \ref{pic2-2}, between the sequences before and after sliding the window, so as to splice back the sequences between different windows.

  \begin{algorithm}  
	\caption{Collision Resolution Protocol}
	\label{alo-1}  
	\begin{algorithmic}[1] 
		\STATE{{\bf Input}: estimated channel $\hat{\bm{H}}$, maximum iteration $t_{max}$, maximum and minimum threshold $\eta$, $\gamma$ }\\
		\STATE{{\bf Output}: recovered channel $\tilde{\bm{H}}$}
		\STATE{{\bf Initialize}: iteration count $t=0$, $\tilde{\bm{H}}=\left[~ \right] $}\\
		\REPEAT
		\STATE{Energy detection:
			\begin{equation}
			\epsilon_{i_k} = \mathbb{E}\left[ \hat{\bm{H}}_{i_k,:} \hat{\bm{H}}_{i_k,:}^H \right] ,i_k \in \left[ 1:2^{B_p}\right] \notag
			\end{equation}
		}\\
		\STATE{Feedback: the BS broadcasts $\left\lbrace i_k, \epsilon_{i_k}>\eta  \right\rbrace $}
		\STATE{Combine: $\tilde{\bm{H}} \leftarrow \left[ \tilde{\bm{H}}; \hat{\bm{H}}_{i_k,:}\right],\left\lbrace i_k\vert \gamma<\epsilon_{i_k}<\eta \right\rbrace $}
		\STATE{Window sliding and retransmission}
		\STATE{Channel estimation: $\hat{\bm{H}}$}
		\STATE{$t \leftarrow t+1$}\\
		\UNTIL{$t=t_{max}$ or $\{\epsilon_{i_k}<\eta, ~\forall i_k \in \left[ 1:2^{B_p} \right] \}$}
		\RETURN{$\tilde{\bm{H}}$}
	\end{algorithmic}  
\end{algorithm} 

\par After obtaining ${\bm{v}_k^p}'$, the CS-based encoder is again performed to encode ${\bm{v}_k^p}'$ to $\bm{a}_{{i_k}'}$ with  ${i_k}'=\left[ {\bm{v}_k^p}' \right]_2 \in \left[ 1: 2^{B_p} \right] $. The encoding process is the same as that in \ref{sec3-1}. Channels of the collided devices are expected to be recovered separately after the retransmission. The ED will be again performed on the recovered channels. If the collision still exists, the window sliding process will be executed again until the maximum number of retransmission is reached or no collision exists. The above collision resolution protocol is summarized in Algorithm \ref{alo-1}. We give the analysis for this collision resolution protocol in Appendix \ref{append-1}, which illustrates that as the window sliding progresses, the number of collided devices will decrease and tend to zero.

 \section{Decoder} \label{sec-5}
 \par The decoding process can be distilled into two key operations: the recovery of the preamble as well as the key parameters for the LDPC code, and the LDPC decoding process combined with SIC \cite{tianya2020gc}. Both are carried out with the MP algorithm. We emphasize again that the beliefs of these two parts can be leveraged to jointly update the messages of the decoding process, which is not considered in \cite{Vem2019gc,Vem2019tcom}. 
 
 \subsection{Joint DAD-CE Algorithm} \label{sec5-1}
 \par The recovery of the preamble as well as the channels in the CS phase is equivalent to the joint DAD-CE problem, which can be modeled as a CS problem. According to the formulation in (\ref{equ-6-add}), the recovery of the sparse matrix $\bm{X}$ can be addressed by the CS-based methods, such as the AMP algorithm and its variants \cite{donoho2009message,Liu2018tsp,ke2020tps,ma2017oamp,GAMP}. Besides, the MP-based approaches \cite{zhang2020iot,hiroki2020arxiv} also work well on the above issues. The Bernoulli and Rician messages are jointly updated in \cite{zhang2020iot}, which considers the fading channel and grant-free scenario. This scenario is also considered in \cite{hiroki2020arxiv}, which takes advantage of estimated data symbols as soft pilot sequences to perform joint channel and data estimation. In this subsection, we consider the MIMO system and derive the update rules of messages based on the BP algorithm.
 
 \par
\begin{figure}[htpb]
	\centerline{\includegraphics[width=0.48\textwidth]{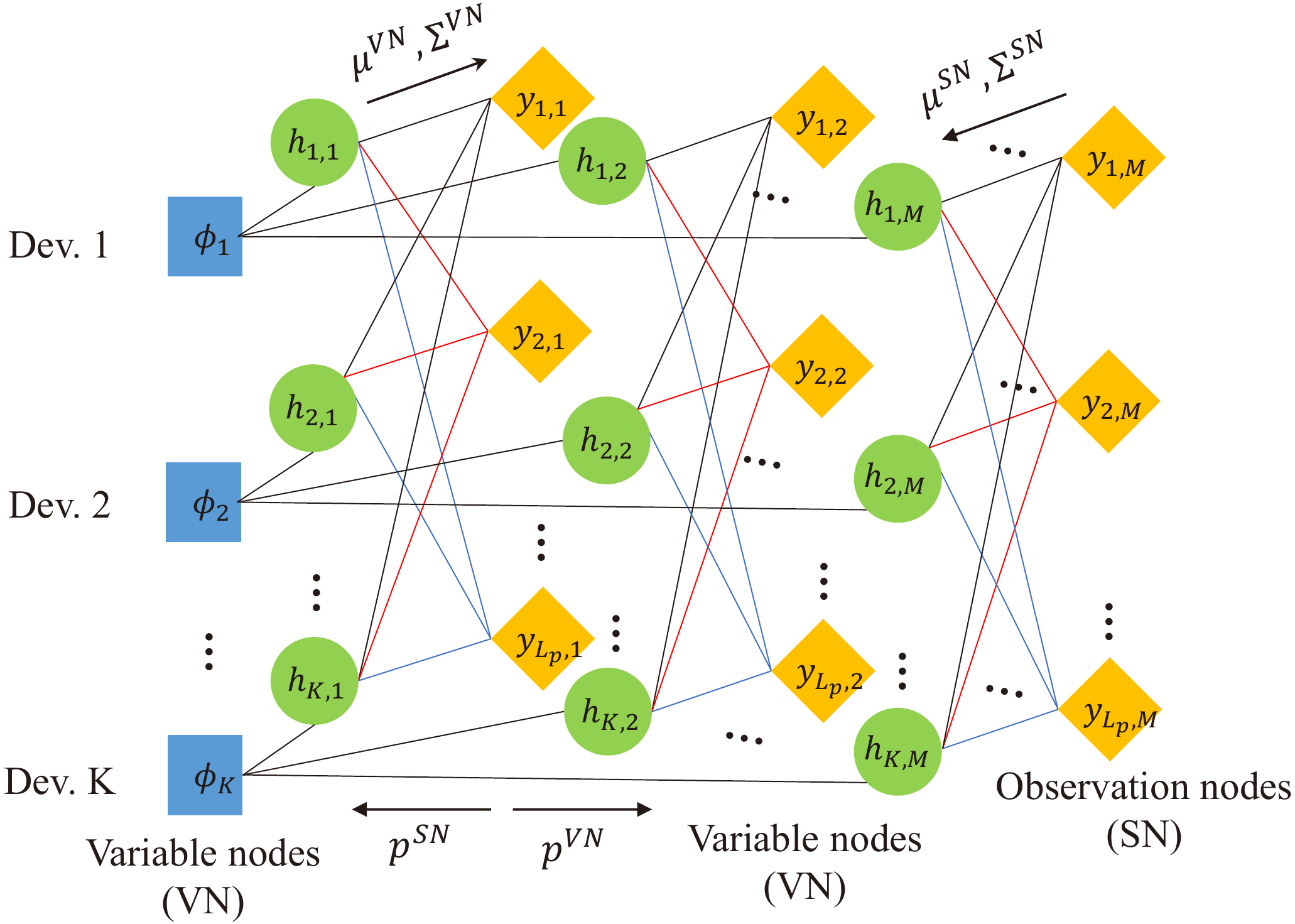}}
 		\caption{Factor graph of the joint DAD-CE algorithm. }
 	\label{pic-3}
 \end{figure}

 We derive the update rules of the activity indicator $\phi_k$ and channel vector $\bm{h}_k$ in (\ref{equ-5}), which are modeled as Bernoulli and Gaussian messages, respectively. The Gaussian messages can be characterized by the estimation of mean-value $\bm{u}_k$ and auto-covariance matrix $\bm{\Sigma}_k$. That is,  $\bm{u}_k$ and $\bm{\Sigma}_k$ are the estimation and estimating deviation of $\bm{h}_k$ in (\ref{equ-5}), respectively. Besides, the Bernoulli messages for the activity indicator can be represented by $p_k$, which is the probability of $\phi_k$ taking the value one. These messages are updated iteratively between the observation and variable nodes, which can be characterized by the factor graph in Fig. \ref{pic-3}, where the received signal $y_{l,m}$ represents the observation node, denoted by SN, and channel $h_{k,m}$ and the activity pattern $\phi_k$ are the variable nodes, denoted by VN. The edges in the factor graph represent the connections among nodes. In the BP algorithm, messages are passed along these edges. We elaborate on the update rules of these messages below.
 
\subsubsection{Message Update at Observation Nodes}
\par We denote $p_{i \rightarrow l}^{VN}(t)$ as the Bernoulli message for the activity of device $i$, which is passed from VN $i$ to SN $l$ in the $t$-th iteration. Accordingly, $\mu_{im \rightarrow lm}^{VN}(t)$ and $\bm{\Sigma}_{i \rightarrow l}^{VN}(t)$ denote the Gaussian messages passed from VN $im$ to SN $lm$, which represent the estimation and the estimating deviation of the channel $\bm{h}_i$, respectively. The index $m \in \left[ 1:M \right] $ denotes the $m$-th antenna and also the $m$-th value of $\bm{h}_i$. Since the update rules of the messages are the same with respect to different iterations, the index of iteration is omitted in the following derivation. For clarity, we assume there is no collision in the CS phase for the following derivation. As such, $i_k$, the original subscript of $\bm{h}$ is replaced by $k$, since there is a one-to-one mapping between these two terms. We emphasize that the collision is considered in our implementation and addressed by the proposed resolution protocol in Algorithm \ref{alo-1}.

\begin{figure}[htpb]
	\centerline{\includegraphics[width=0.45\textwidth]{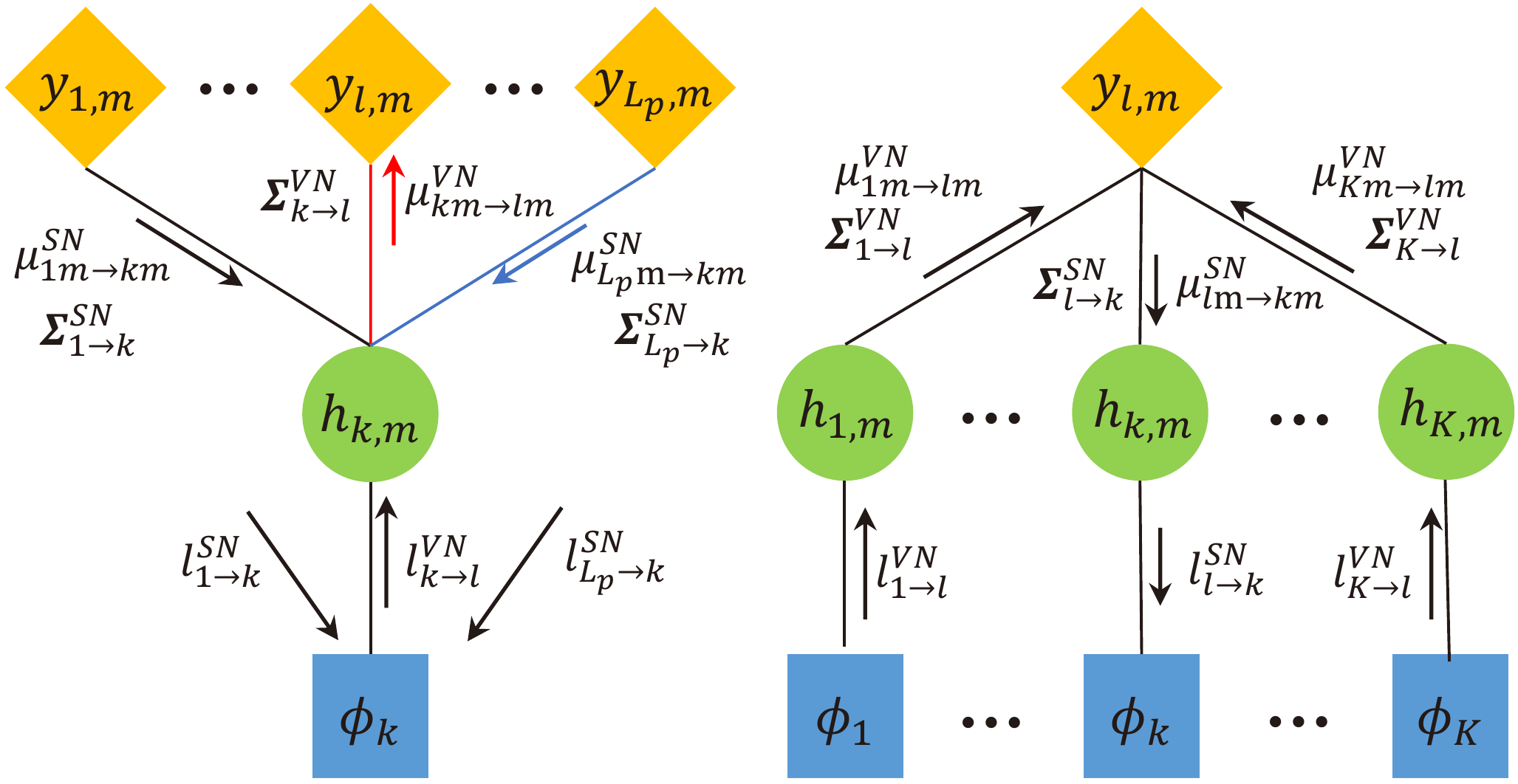}}
	\caption{Message update rules at VNs and SNs. The output message on each edge is obtained by collecting the messages from the other edges connected with the same node. }
	\label{pic-3-add}
\end{figure}

To give the message update rules at the SN $y_{l,m}$ in Fig. \ref{pic-3-add}, we first rewrite \eqref{equ-5} as
\begin{equation}\label{equ-13}
	\begin{aligned}
		y_{lm}&= \sum\nolimits_{i=1}^{K}{A_{li}\phi_ih_{im}} + n_{lm} \\
		&= A_{lk}\phi_kh_{km} + \underbrace{\sum\nolimits_{i\in \mathcal{K} \backslash k}{A_{li}\phi_ih_{im}} + n_{lm}}_{z_{lkm}},
	\end{aligned}
\end{equation}
where $l$ is in $L_p$ and $ \mathcal{K} \backslash k$ denotes the entries in set $\left\lbrace 1,2,\cdots,K\right\rbrace $ except $k$. The term $\sum\nolimits_{i\in K \backslash k}{A_{li}\phi_ih_{im}} + n_{lm}$ is modeled as an equivalent Gaussian noise with $\bm{z}_{lk}\sim \mathcal{CN}(\bm{\mu}_{z_{lk}},\bm{\Sigma}_{z_{lk}})$, where $\bm{\mu}_{z_{lk}}=[\mu_{z_{lk1}},\cdots,\mu_{z_{lkm}},\cdots,\mu_{z_{lkM}}]$ and $\mu_{z_{lkm}}$ is given by
\begin{equation}
	\mu_{z_{lkm}}= \sum\nolimits_{i \in \mathcal{K} \backslash k} {A_{li} \cdot p_{i \rightarrow l}^{VN} \cdot \mu_{im \rightarrow lm}^{VN}}. 
	\label{equ-14}
\end{equation}
For $\bm{\Sigma}_{z_{lk}}$, the auto-covariance of $\bm{z}_{lk}$, since $\phi_i$ is the same for $M$ antennas, resulting in the correlation among antennas, we should not consider the variance of $\bm{z}_{lkm}$ at each antenna $m$ separately. Instead, the covariance of $z_{lk}$ is considered in this paper. Rewrite (\ref{equ-13}) in a vector form:
\begin{equation}
	\begin{aligned}
			\mathbf{y}_l^T &= \left[ {\begin{array}{*{20}{c}}
				y_{l1}\\
				y_{l2}\\
				\vdots \\
				y_{lM}
		\end{array}} \right] = A_{lk}\phi_k \cdot \left[ {\begin{array}{*{20}{c}}
				h_{k1}\\
				h_{k2}\\
				\vdots \\
				h_{kM}
		\end{array}} \right] \\
	&+\underbrace{\sum\nolimits_{i \in \mathcal{K} \backslash k}{A_{li}\phi_i\cdot \left[ {\begin{array}{*{20}{c}}
						h_{i1}\\
						h_{i2}\\
						\vdots \\
						h_{iM}
				\end{array}} \right]} + \left[ {\begin{array}{*{20}{c}}
					n_{l1}\\
					n_{l2}\\
					\vdots \\
					n_{lM}
			\end{array}} \right]}_{\mathbf{z}_{lk}}.
	\end{aligned} \label{equ-15}
\end{equation}
The $(m,n)$-th $(m\neq n)$ entry for $\bm{\Sigma_{z_{lk}}}$ satisfies 
\begin{equation}
	\begin{aligned}
		(\Sigma_{z_{lk}})_{(m,n)} & = \sum\nolimits_{i \in \mathcal{K} \backslash k}{{\left| {{A_{li}}} \right|}^2\cdot p_{i \to l}^{VN}\cdot\Big\{ (\Sigma_{i \to l}^{VN})_{(m,n)} } \\
	&{+ ~q_{i \to l}^{VN} \cdot \mu_{im \to lm}^{VN}\cdot(\mu_{in \to ln }^{VN})^* \Big\}}, ~m \neq n,
	\end{aligned}
\label{equ-16}
\end{equation}
where $q_{i \to l}^{VN} =  1 - p_{i \to l}^{VN}$ denotes the probability that the Bernoulli variable $\phi_k$ equals to zero. If $m=n$, we have 
	\begin{equation}
		\begin{aligned}
			(\Sigma_{z_{lk}})_{(m,m)} &= \sum\nolimits_{i \in \mathcal{K} \backslash k}{{\left| {{A_{li}}} \right|}^2\cdot p_{i \to l}^{VN}\cdot\Big\{(\Sigma_{i \to l}^{VN})_{(m,m)}}\\ 
				&{\left.\quad \quad \quad +~q_{i \to l}^{VN} \cdot \left| \mu_{im \to lm}^{VN}\right| ^2 \right\rbrace} + \sigma_n^2.			
		\end{aligned}
\label{equ-17}
\end{equation}

\par Details about the derivation of $\bm{\Sigma_{z_{lk}}}$ are given in Appendix \ref{append-2-1}. After obtaining the mean and covariance of $\bm{z}_{lk}$, we can get the Gaussian messages $\mu_{lm \rightarrow km}^{SN}$ and $\bm{\Sigma}_{l \rightarrow k}^{SN}$ passed from SN $lm$ to VN $km$ as below 
\begin{align}
	\begin{split}
		\mu_{lm \to km}^{SN} &= \mathbb{E} \left[ h_{km} \vert y_{lm}, \mu_{z_{lkm}}, \bm{\Sigma}_{z_{lk}}, \phi_k=1 \right] \\
		&= (y_{lm}-\mu_{z_{lkm}}) \slash A_{lk} \label{equ-18}
	\end{split}\\
	\begin{split}
		\bm{\Sigma}_{l \to k}^{SN}&= \text{Var} \left[ h_{km} \vert y_{lm}, \mu_{z_{lkm}}, \bm{\Sigma}_{z_{lk}}, \phi_k=1 \right] \\
		&= \bm{\Sigma}_{z_{lk}} \slash \left| A_{lk} \right| ^2,\label{equ-19}
	\end{split}
\end{align}
where $\mathbb{E}\left[ a|b\right] $ and $\text{Var}\left[ a|b\right] $ denote the expectation and variance of $a$ conditioned on $b$, respectively.

\par For the Bernoulli message $p_{l \to k}^{SN}$ passed for SN $l$ to VN $k$,  we have 
\begin{equation}
	\begin{aligned}
		p_{l \to k}^{SN} &= \left[ 1+ \frac{P(\mathbf{y}_l \vert \phi_k=0, \bm{\mu}_{z_{lk}},\bm{\Sigma}_{z_{lk}})}{P(\mathbf{y}_l \vert \phi_k=1, \bm{\mu}_{z_{lk}},\bm{\Sigma}_{z_{lk}})}\right] ^{-1} \\
		&= \left[ 1+ \frac{P(\bm{y}_l=\bm{z}_{lk} \vert \bm{\mu}_{z_{lk}}, \bm{\Sigma}_{z_{lk}})}{P(\bm{y}_l= A_{lk}\cdot\bm{h}_k+ \bm{z}_{lk} \vert \bm{\mu}_{z_{lk}}, \bm{\Sigma}_{z_{lk}})} \right] ^{-1} \\
		&= \left[ 1+ \frac{f(\bm{y}_l\vert \bm{\mu}_{z_{lk}}, \bm{\Sigma}_{z_{lk}})}{f(\bm{y}_l\vert \bm{\mu}_{z_{lk}}^{'}, \bm{\Sigma}_{z_{lk}}^{'})} \right] ^{-1} \\
		&= \frac{f(\bm{y}_l\vert \bm{\mu}_{z_{lk}}^{'}, \bm{\Sigma}_{z_{lk}}^{'})}{f(\bm{y}_l\vert \bm{\mu}_{z_{lk}}, \bm{\Sigma}_{z_{lk}})+f(\bm{y}_l\vert \bm{\mu}_{z_{lk}}^{'}, \bm{\Sigma}_{z_{lk}}^{'})},
	\end{aligned} \label{equ-20}
\end{equation}
where 
\begin{align}
	\bm{\mu}_{z_{lk}}^{'} &= A_{lk} \cdot \bm{\mu}_{k \to l}^{VN} + \bm{\mu}_{z_{lk}} \\
	\bm{\Sigma}_{z_{lk}}^{'}   &= \left| A_{lk} \right| ^2 \cdot \bm{\Sigma}_{k \to l}^{VN} + \bm{\Sigma}_{z_{lk}},
\end{align}
which denote the mean-value and covariance of $\bm{y}_l$ when $\phi_k=1$, respectively. And $f(\bm{x} \vert \bm{\mu}, \Sigma)$ denotes the probability density function (pdf)  of the multi-dimensional complex Gaussian distribution $\mathcal{CN}_M(\bm{x} \vert \bm{\mu}, \Sigma)$, that is
\begin{equation}
	f(\bm{x} \vert \bm{\mu}, \bm{\Sigma})= \frac{1}{\pi^M \!\cdot\! \det\left(  \bm{\Sigma}\right)  }\!\cdot \exp \left[ - (\bm{x}-\bm{\mu})^H \bm{\Sigma}^{-1}  (\bm{x}-\bm{\mu})\right]. \label{equ-23}
\end{equation}
Moreover, the Bernoulli message can be simplified by the use of log-likelihood ratio (LLR) to reduce the complexity as well as to avoid the computation overflow. Hence, the LLR of the message in (\ref{equ-20}) can be represented as
\begin{equation}
	\begin{aligned}
		l_{l \to k}^{SN} & \triangleq \text{ln}\frac{P(\phi_{k}=1)}{P(\phi_{k}=0)} = \ln \frac{p_{l \to k}^{SN}}{1-p_{l \to k}^{SN}} \\
		& \overset{(a)}{=} \ln \frac{\det\left( \bm{\Sigma}_{z_{lk}}\right)  }{\det\left(  \bm{\Sigma}_{z_{lk}}^{'}\right) } \!+\! (\bm{y}_l-\bm{\mu}_{z_{lk}})^H\cdot \bm{\Sigma}_{z_{lk}}^{-1} \cdot (\bm{y}_l-\bm{\mu}_{z_{lk}}) \\
		& \quad \quad \quad -  (\bm{y}_l-\bm{\mu}_{z_{lk}}^{'})^H\cdot (\bm{\Sigma}_{z_{lk}}^{'})^{-1} \cdot (\bm{y}_l-\bm{\mu}_{z_{lk}}^{'}),
	\end{aligned} \label{equ-24}
\end{equation}
where $\overset{(a)}{=}$ is derived by the substitution of (\ref{equ-20}) and (\ref{equ-23}). 

\subsubsection{Message Update at Variable Nodes}
\par Likewise, the Gaussian and Bernoulli messages at VNs are updated by collecting the incoming messages from SNs. To ensure convergence, messages from the VN's own are not included in the calculation \cite{BP}. Typically, for the Gaussian messages, since $\bm{h}$ follows the Gaussian distribution, the update rule at the VN is to multiply the pdfs observed at each SN to obtain a new one. We note that the prior Gaussian distribution of $\bm{h}$ is also included in the multiplication. The new pdf still follows the Gaussian distribution, of which the mean and covariance are the updated messages at the VN. The pdf of $\bm{h}_k$ passed from VN $k$ to SN $l$ is given by
\begin{equation}
	\begin{aligned}
		&f(\bm{h} \vert \bm{\mu}_{k \to l}^{VN}, \bm{\Sigma}_{k \to l}^{VN}) \\ 
		& \quad \quad \propto \prod\nolimits_{i \in \mathcal{L} \backslash l} {f(\bm{h} \vert \bm{\mu}_{i \to k}^{SN}, \bm{\Sigma}_{i \to k}^{SN})} \cdot f(\bm{h} \vert \bm{\mu}_k^{pri}, \bm{\Sigma}_k^{pri}). \label{equ-25}
	\end{aligned}
\end{equation}
Accordingly, for the Gaussian pdf $f(\bm{h} \vert \bm{\mu}_{k \to l}^{VN}, \bm{\Sigma}_{k \to l}^{VN})$, the mean and covariance are give by
\begin{align}
	\begin{split}
		\bm{\mu}_{k \to l}^{VN} &= \mathbb{E}\left[ \bm{h}_k \vert \bm{\mu}_{i \to k}^{SN}, \bm{\Sigma}_{i \to k}^{SN}, i\in \mathcal{L} \backslash l \right]  \\											   
		&= \bm{\Sigma}_{k \to l}^{VN} \cdot \Big[ (\bm{\Sigma}_{k}^{pri})^{-1}\cdot \bm{\mu}_k^{pri} \\
		& \quad \quad \quad \quad +  \sum\nolimits_{i \in \mathcal{L} \backslash l} {(\bm{\Sigma}_{i \to k}^{SN})^{-1} \cdot \bm{\mu}_{i \to k}^{SN}} \Big] \label{equ-26}
	\end{split}\\
	\begin{split}
		\bm{\Sigma}_{k \to l}^{VN} 	&= \operatorname{Var}\left[ \bm{h}_k \vert \bm{\mu}_{i \to k}^{SN}, \bm{\Sigma}_{i \to k}^{SN}, i\in \mathcal{L} \backslash l \right]  \\
		&= \left[ \sum\nolimits_{i \in \mathcal{L} \backslash l}{(\bm{\Sigma}_{i \to k}^{SN})^{-1}} + (\bm{\Sigma}_k^{pri})^{-1} \right] ^{-1}, \label{equ-27}
	\end{split}				
\end{align}
where $\bm{\mu}_k^{pri}= \bm{0}_{M \times 1}$ and $ \bm{\Sigma}_k^{pri}= \bm{I}_M$ are the prior mean and covariance of $\bm{h}_k$, $ \mathcal{L} \backslash l$ denotes the entries in set $\left\lbrace 1,2,\cdots,L\right\rbrace $ except $l$. We give the derivations of (\ref{equ-26}) and (\ref{equ-27}) in Appendix \ref{append-2-2}.

\par The derivation of the Bernoulli messages is the same as above, which is updated by collecting the messages observed at SNs. For $p_{k \to l}^{VN}$ passed for VN $k$ to SN $l$, it is obtained by multiplying the probability of $\phi_k=1$ passed from all the SNs to VN $k$ and then normalizing. Likewise, for convergence, the message passed from SN $l$ to VN $k$ is not included. We emphasize that the prior activation probability of each device $p_a$ is also considered. As such, $p_{k \to l}^{VN}$ is given by 
\begin{equation}
	\begin{aligned}
		p_{k \to l}^{VN} &= P\left( \phi_k=1 \vert  \left\lbrace p_{i\to k}^{SN}, i\in \mathcal{L} \backslash l \right\rbrace, p_a \right) \\
		&= \frac {p_{a}\cdot \prod _{i\in \mathcal {L}\backslash l}p^{SN}_{i\to k}}{p_{a} \cdot \prod _{i\in \mathcal {L} \backslash l}p^{SN}_{i\to k}+(1-p_{a})\cdot \prod _{i\in \mathcal {L}\backslash l}\left ({1-p^{SN}_{i\to k}}\right)}. 
	\end{aligned} 
\end{equation}
Likewise, for complexity reduction, we also employ the LLR to represent this message in iterations, of which the relationship with the activation probability is 
\begin{equation}
	\begin{aligned}
			l_{k \to l}^{VN}  &= \ln \frac {p^{VN}_{k\to l}}{1-p^{VN}_{k\to l}}=l_{0}+\sum _{i\in \mathcal {L}\backslash l}l^{SN}_{i\to k}\\
			p^{VN}_{k\to l} &=\frac {1}{1+\exp \left ({{-l^{VN}_{k\to l}}}\right)}, \label{equ-29}
	\end{aligned}
\end{equation}
where $l_0= \ln \frac{p_a}{1-p_a}$ is the prior LLR of the probability for the device being active.

\subsubsection{DAD Decision and CE Output}
\par Since the messages above are iteratively updated between SNs and VNs, after reaching the maximum number of iterations, the Bernoulli and Gaussian messages will have an output at VNs. For the Gaussian messages, similar to the above update rules, the output is obtained by combining all the incoming messages from SNs, i.e,
\begin{align}
	\begin{split}
		\bm{\mu}_{k}^{dec} &= \bm{\Sigma}_{k}^{dec} \cdot \left[ (\bm{\Sigma}_{k}^{pri})^{-1}\cdot \bm{\mu}_k^{pri} \right. \\
		& \quad \quad \quad \quad \left.+  \sum\nolimits_{i \in \mathcal{L}} {(\bm{\Sigma}_{i \to k}^{SN})^{-1} \cdot \bm{\mu}_{i \to k}^{SN}} \right] \label{equ-30}
	\end{split}\\
	\begin{split}
		\bm{\Sigma}_{k}^{dec} 	&= \left[ \sum\nolimits_{i \in \mathcal{L}}{(\bm{\Sigma}_{i \to k}^{SN})^{-1}} + (\bm{\Sigma}_k^{pri})^{-1} \right] ^{-1}, \label{equ-31}
	\end{split}				
\end{align} 
which denote the output estimation and estimating deviation of $\bm{h}_k$, respectively. For the Bernoulli messages, the LLR of the DAD decision is 
\begin{equation}
	l^{\text {dec}}_{k}=l_{0}+\sum\nolimits_{l\in \mathcal {L}}l^{SN}_{l\to k}+l^{ce}_{k}. \label{equ-32}
\end{equation}
Device $k$'s activity is detected as $\hat{\phi}_k=1$ if $ l^{\text {dec}}_{k}>0$ and vice versa. The term $l_k^{ce}$ in (\ref{equ-32}) is to improve the DAD accuracy by exploiting the CE result \cite{zhang2020iot}, which is derived as follows. The estimated channel $\bm{h}_k$ can be modeled as $\hat{\bm{h}}_k= \bm{h}_k + \bm{\epsilon}_k$, where $\bm{\epsilon}_k$ is the complex Gaussian noise distributed as $\bm{\epsilon}_k \sim \mathcal{CN}(0,\bm{\Sigma}_{k}^{dec})$. Accordingly, the distribution of $\hat{\bm{h}}_k$ with respect to value of $\phi_k$ is
\begin{equation}
\hat{\bm{h}}_k	\sim \left\{\begin{array}{l c}
		\mathcal{CN}(\bm{\mu}_k^{pri},\bm{\Sigma}_{k}^{pri}+\bm{\Sigma}_{k}^{dec}), &\phi_k=1 \\
		\mathcal{CN}(\bm{0},\bm{\Sigma}_{k}^{dec}), &\phi_k=0
	\end{array} \quad \forall k \in \mathcal{K}_{tot}.\right. \label{equ-33}
\end{equation}
Therefore, this information can be leveraged to give an extra belief to the DAD decision. Similar to (\ref{equ-20}), $l^{ce}_{k}$ is given by

\begin{equation}
	\begin{aligned}
			l^{ce}_{k} =&\ln \frac {P\left ({\hat {\bm{h}}_{k} =\bm{\mu}_{k}^{dec} \vert \phi _{k}=1, \bm{\mu}_{k}^{pri},\bm{\Sigma}_{k}^{dec},\bm{\Sigma}_{k}^{pri} }\right)}{P\left ({\hat {\bm{h}}_{k} =\bm{\mu}_{k}^{dec} \vert \phi _{k}=0,\bm{\Sigma}_{k}^{dec} }\right)} \\
		=&\ln \frac {f\left ({\bm{\mu}_{k}^{dec} \vert \bm{\mu}_{k}^{pri},\bm{\Sigma}_{k}^{pri}+\bm{\Sigma}_{k}^{dec}}\right)}{f\left ({\bm{\mu}_{k}^{dec} \vert \bm{0},\bm{\Sigma}_{k}^{dec}}\right)} \\
		=&\ln \frac {\det\left( \bm{\Sigma}_{k}^{dec}\right) }{\det\left( \bm{\Sigma}_{k}^{pri}\!+\!\bm{\Sigma}_{k}^{dec}\right) } + \left(\bm{\mu}_k^{dec}\right)^H\cdot \left(\bm{\Sigma}_{k}^{dec} \right) ^{-1} \cdot \bm{\mu}_k^{dec} \\
		&- \left( \bm{\mu}_k^{dec} \!-\!\bm{\mu}_k^{pri}\right) ^H\ \!\!\!\cdot \! \left(\bm{\Sigma}_{k}^{pri} \!+\! \bm{\Sigma}_{k}^{dec} \right) ^{-1} \!\! \!\cdot \! \left( \bm{\mu}_k^{dec}\!-\!\bm{\mu}_k^{pri}\right) \!. \label{equ-34}
	\end{aligned}
\end{equation}
As aforementioned, we derive the LLR expression by utilizing the joint distribution of the channel among antennas. Finally, we obtain the estimated channel of device $k$ as 
\begin{equation}
	\bm{\hat{h}}_k = \hat{\phi}_k \cdot \bm{\mu}_{k}^{dec}.
\end{equation}
\begin{figure}[htpb]
	\centerline{\includegraphics[width=0.4\textwidth]{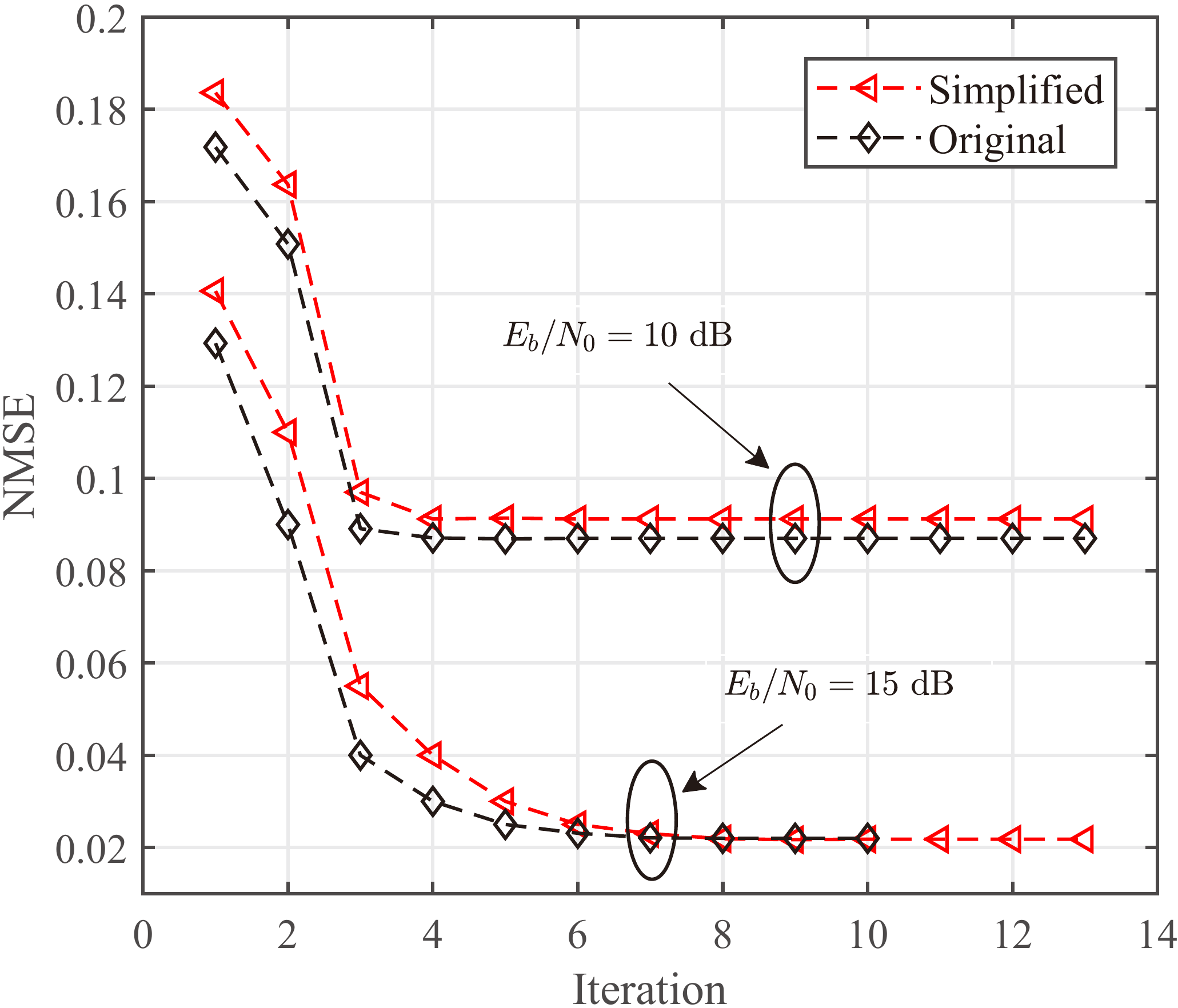}}
	\caption{NMSE of CE by the Joint DAD-CE algorithm. $B_p=9$, $K_a=30$, $M=30$, $L_p=200$.}
	\label{pic-5-add}
\end{figure}

\par The above joint DAD-CE algorithm is summarized in Algorithm \ref{alo-2}, where $N_{iter}$ denotes the maximum number of iterations. We note that $\bm{\Sigma}^{VN}$ and $\bm{\Sigma}^{SN}$ both go to diagonal matrices over the iteration in our numerical verification. Hence, the corresponding matrix inverse operations can be simplified to the divisions to reduce the complexity with little performance loss. As shown in Fig. \ref{pic-5-add}, Simplified denotes the approximation by treating $\bm{\Sigma}^{VN}$ and $\bm{\Sigma}^{SN}$ as diagonal matrices and Original means there is no approximation in Algorithm \ref{alo-2}. We use normalized mean square error (NMSE) for the evaluation of the CE performance. Fig. \ref{pic-5-add} illustrates that this approximation introduces little performance loss, which, although, has greatly reduced the complexity as aforementioned.

  \begin{algorithm}  
	\caption{Joint DAD-CE Algorithm}
	\label{alo-2}  
	\begin{algorithmic}[1]	
		\STATE {{\bf Input}: $\bm{Y_p}$, $\bm{A}$, $\bm{\mu}^{pri}$, $\bm{\Sigma}^{pri}$, $\sigma_n^2$, $p_a$}\\
		\STATE{{\bf Initialize}: $\bm{\mu}^{VN}=\bm{\mu}^{pri}, \bm{\Sigma}^{VN}=\bm{\Sigma}^{pri}$}
		\REPEAT
			\STATE {SN update: $\bm{\mu}^{SN}$, $\bm{\Sigma}^{SN}$ by \eqref{equ-18}-\eqref{equ-19}}\\
			\STATE {SN update: $l^{SN}$ by \eqref{equ-24}}\\
			\STATE {VN update: $\bm{\mu}^{VN}$, $\bm{\Sigma}^{VN}$ by \eqref{equ-26}-\eqref{equ-27}}\\
			\STATE {VN update: $l^{VN}$ by \eqref{equ-29}}\\
		\UNTIL {$N_{iter}$ reached}
		\STATE {CE output: $\bm{\mu}^{dec}$, $\bm{\Sigma}^{dec}$ by \eqref{equ-30}-\eqref{equ-31}}\\
		\STATE {DAD Decision: $l^{dec}$ by \eqref{equ-32}}\\
		\RETURN {$\left\lbrace\hat{\bm{h}}_k, \hat{\phi}_k, \forall k \in \mathcal{K}_{tot} \right\rbrace $}
	\end{algorithmic}  
\end{algorithm} 

\begin{figure*}[t]
	\centerline{\includegraphics[width=1.0\textwidth]{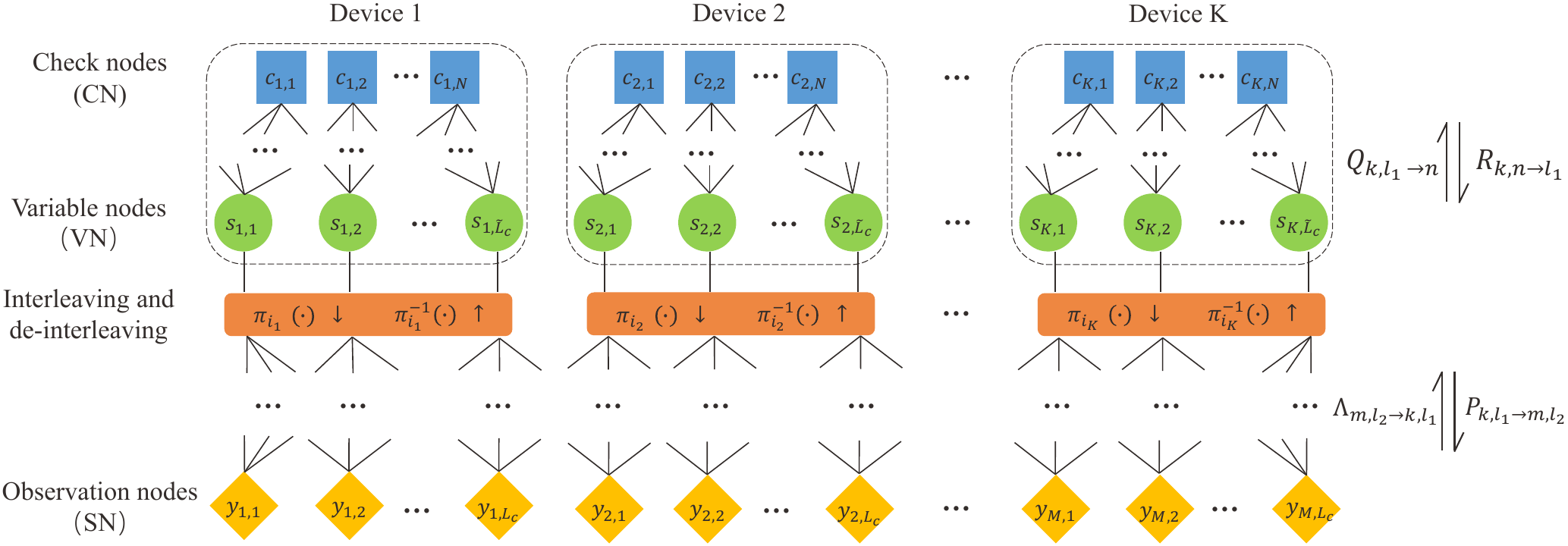}}
	\caption{Factor graph for LDPC decoding. }
	\label{pic-4}
\end{figure*}

 \subsection{MIMO-LDPC-SIC Decoder}
 \par After obtaining the key parameters, such as interleaving patterns and channels by the joint DAD-CE algorithm, the LDPC decoding problem can be addressed by the standard BP algorithm \cite{BP,LDPC-MIMO}. Likewise, the DD process is performed by updating messages iteratively at different nodes. Differently, since the LDPC is a forward error correction code, besides the observation and variable nodes, the check nodes (CNs) are considered in the factor graph to provide an extra belief, as shown in {Fig. \ref{pic-4}}. We rewrite the received signal in the LDPC phase as 
\begin{equation}
	{\mathbf{Y}_c} = {\mathbf{Y}_{{L_p} + 1:{L},:}} = \sum\limits_{k \in {\mathcal{K}_a}} {{\pi _{{i_k}}}\left( {{\tilde{\mathbf{s}}_k}} \right){\mathbf{h}_{{i_k}}^T} + {\mathbf{Z}_{_{{L_p} + 1:{L},:}}}},
	\label{equ-35}
\end{equation}
where $\mathbf{Y}_c \in {C^{{L_c} \times M}}$ is the last $L_c$ rows of $\mathbf{Y}$. The LDPC decoder is tasked to recover the last $B_c$ bits of information based on the received signal $\mathbf{Y_c}$, estimated interleaving patterns and channels using the low-complexity iterative BP algorithm. Owing to the two-phase encoding scheme, these key parameters can be recovered in the decoding of the CS phase. We emphasize that once the active indicators $\phi_k$ are recovered, the positions of $\left\lbrace \phi_k=1, k\in \mathcal{K}_{tot}\right\rbrace $ in the selection matrix $\bm{\Phi}$, i.e., $\left\lbrace i_k, \forall k\in \mathcal{K}_a \right\rbrace $, are determined. Thereby, the interleaving patterns of active devices are also recovered. As shown in \eqref{equ-35}, the zero-padded sequences $\tilde{\mathbf{s}}_k$ are subject to different permutations, and each is determined by the interleaving pattern $i_k$. Therefore, the effect of these permutations needs to be considered when the messages are being sent to and from the VNs, i.e., interleaving and de-interleaving in Fig. \ref{pic-4}, respectively.

\par The connections among nodes in Fig. \ref{pic-4} appear to be more involved than those of Fig. \ref{pic-3}. In the upper part of Fig. \ref{pic-4}, the CNs (blue color) and VNs (green color) as well as the edges connecting them constitute the Tanner graph in LDPC. The subscript $N=\tilde{L}_c-B_c$ denotes the number of CNs in the LDPC code, which corresponds to the number of rows of the LDPC check matrix. $K$ denotes the number of active devices estimated in the CS phase. Other subscripts are consistent with the aforementioned. The edges between CNs and VNs are described by the LDPC check matrix, which cannot be marked explicitly in the graph. For example, in the check matrix of device $k$, if the entry $c_{i,j}=1$, there will be an edge between SN $c_{k,i}$ and VN $s_{k,j}$.

\begin{figure}[htpb]
	\centerline{\includegraphics[width=0.3\textwidth]{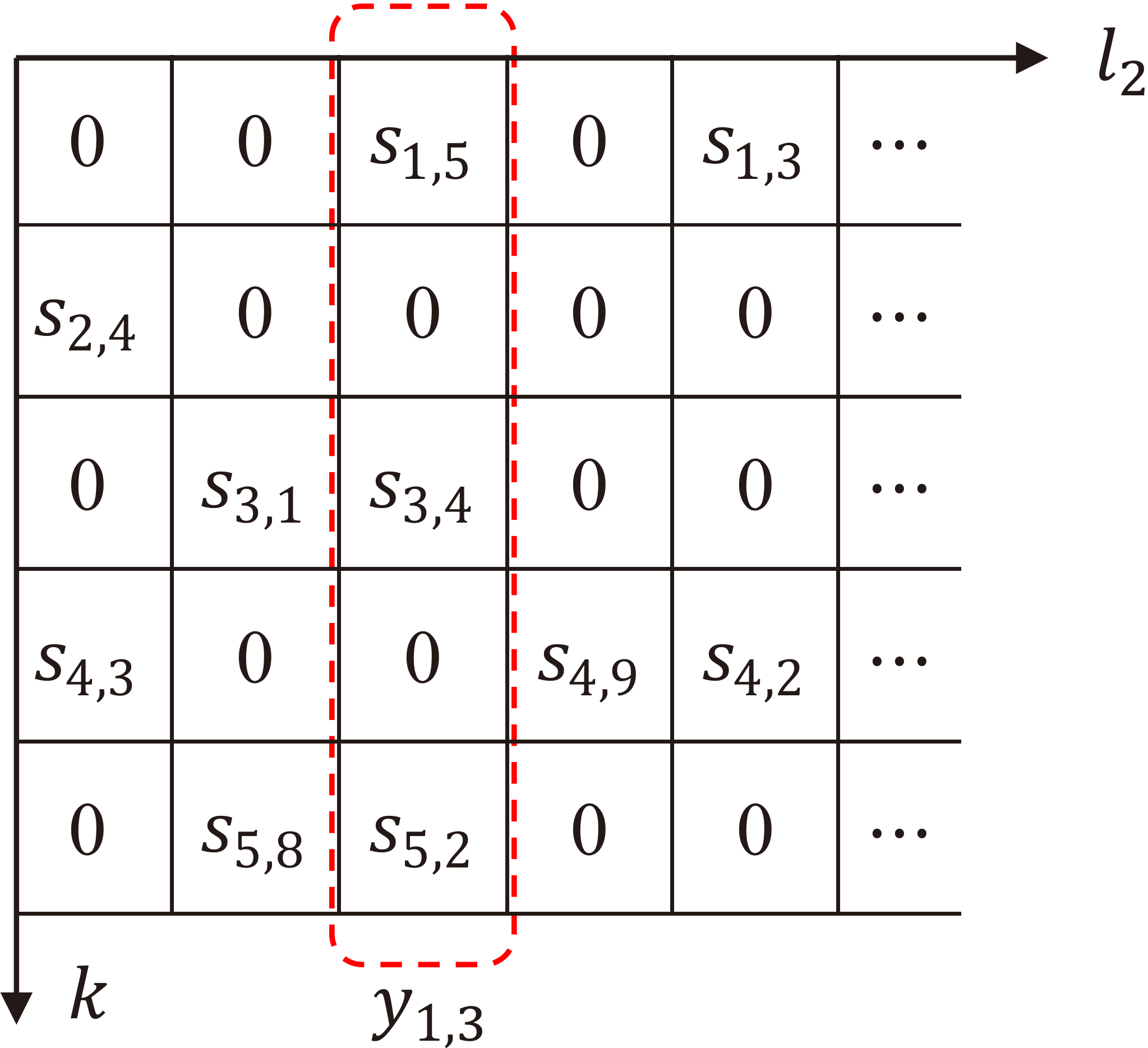}}
	\caption{The set of VNs connected with an SN is decided by the interleaving patterns. For instance, in the diagram, the set of VNs related with $y_{1,3}$ is $\left\lbrace s_{1,5}, s_{3,4}, s_{5,2} \right\rbrace $. Correspondingly, the set of SNs connected to these VNs is $\left\lbrace y_{m,3}, m=1,2,\cdots,M \right\rbrace $.}
	\label{pic-7}
\end{figure}

\par The lower part of Fig. \ref{pic-4} refers to the graph for MIMO detection, of which the edges between VNs and SNs (yellow color) are simply determined by \eqref{equ-35} though looking complicated. For example, the VN $s_{k,l_1} \left( l_1\in \left[1:\tilde{L}_c \right]\right)  $ is connected to the $l_2$-th SN from all antennas (i.e., $y_{m,l_2},m=1,2,...,M,l_2\in \left[1:L_c \right] $). We note that $l_1$ is not necessarily equal to $l_2$ in the presence of zero-padding and interleaving. Correspondingly, the VNs connected to SN $y_{m,l_2}$ depend on whose data is interleaved to the $l_2$-th channel use. For instance, as illustrated in Fig. \ref{pic-7}, the set of VNs connected to SN $y_{1,3}$ is $\left\lbrace s_{1,5}, s_{3,4}, s_{5,2} \right\rbrace $. That is, after zero-padding and interleaving, the fifth, fourth, and second bits of devices $1$, $3$, and $5$ are mapped to the third channel use, respectively. Before conducting the MP algorithm, we define the types of messages as follows.

\begin{itemize}
	\item ${R_{k,n \to l_1}}$: Messages passed from CN $c_{k,n}$ to VN $s_{k,l_1}$.
	\item ${Q_{k,l_1 \to n}}$: Messages passed from VN $s_{k,l_1}$ to CN $c_{k,n}$.
	\item ${P_{k,l_1 \to m,l_2}}$: Messages passed from VN $s_{k,l_1}$ to SN $y_{m,l_2}$.
	\item ${\Lambda _{m,l_2 \to k,l_1}}$: Messages passed from SN $y_{m,l_2}$ to VN $s_{k,l_1}$.
\end{itemize}

\par The massages $R$ and $Q$ refer to the parity check constraints in the LDPC code, while $P$ and $\Lambda$ are related to the received signals in the MIMO system. 

\begin{figure}[htpb]
	\centerline{\includegraphics[width=0.5\textwidth]{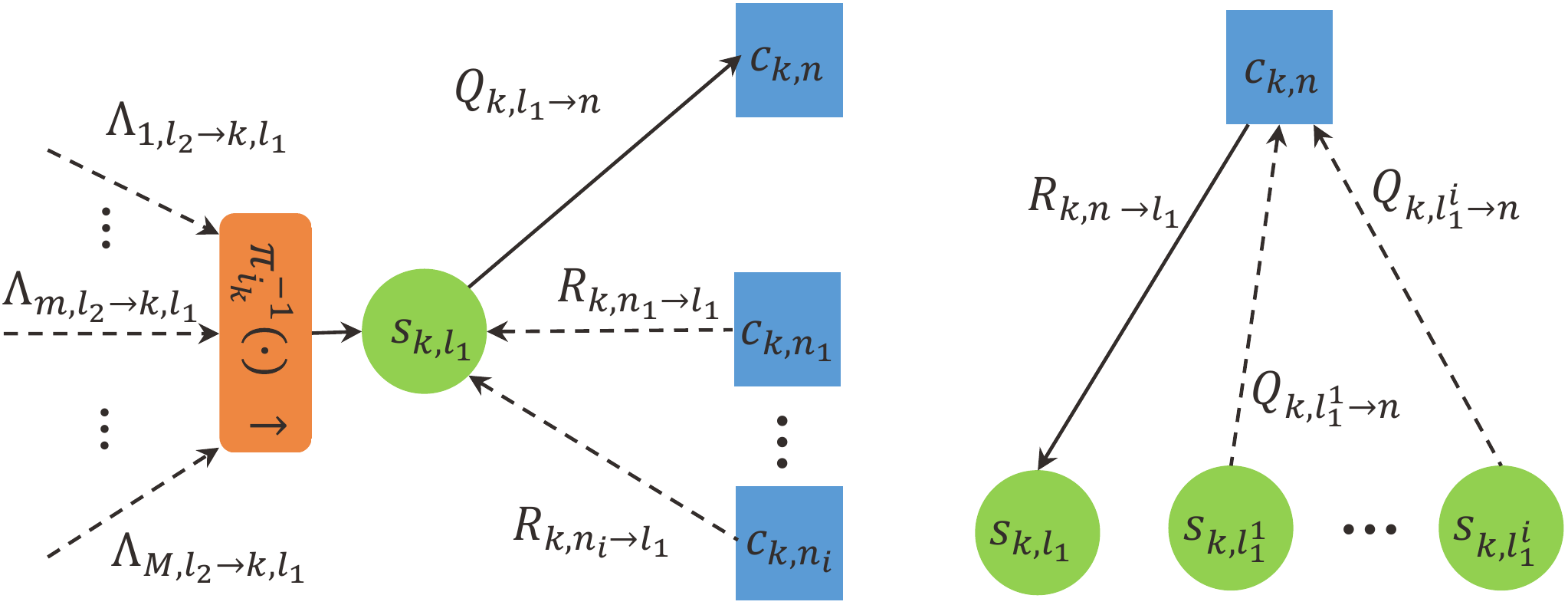}}
	\caption{Update rules for messages $R$ and $Q$ at CNs and VNs on the Tanner graph for device $k$. The dashed and solid lines represent input and output messages, respectively.}
	\label{pic-5}
\end{figure}

\par We first give the MP rules for the LDPC decoding with BPSK modulation, which is known as the sum-product algorithm. As illustrated in Fig. \ref{pic-5}, the messages $R$ and $Q$ are iteratively updated between CNs and VNs. Likewise, for the reduction of complexity, we give the updating rules in the LLR form.
\begin{align}
	{Q_{k,l_1 \to n}} &= \sum\limits_{j \in M} {\pi_{i_j}^{-1}\left( \Lambda _{j,l_2 \to k,l_1}\right) } \! + \!\!  \sum\limits_{j \in {\mathcal{N}_c}\left( {k,l_1} \right) \backslash n} {\!\!\!{R_{k,j \to l_1}}}  \label{equ-36}\\
	{R_{k,n \to l_1}} &= 2{\tanh ^{ - 1}}\left( {\prod\limits_{j \in {\mathcal{N}_v}\left( {k,n} \right) \backslash l_1} {\!\!\!\tanh \left( {\frac{{{Q_{k,j \to n}}}}{2}} \right)} } \right),	\label{equ-37}
\end{align}
where ${\mathcal{N}_c}\left( {k,l_1} \right) \backslash n$ denotes the set of CNs connected to $s_{k,l_1}$ except $c_{k,n}$, i.e., $\left\lbrace c_{k,n},c_{k,n_1},\cdots, c_{k,n_i}\right\rbrace $ in Fig. \ref{pic-5}. Likewise, ${\mathcal{N}_v}\left( {k,n} \right) \backslash l_1$ denotes the set of VNs connected to $c_{k,n}$ except $s_{k,l_1}$, i.e., $\left\lbrace s_{k,l_1^1},\cdots,s_{k,l_1^i}\right\rbrace $ in Fig. \ref{pic-5}.

\begin{figure}[htpb]
	\centerline{\includegraphics[width=0.5\textwidth]{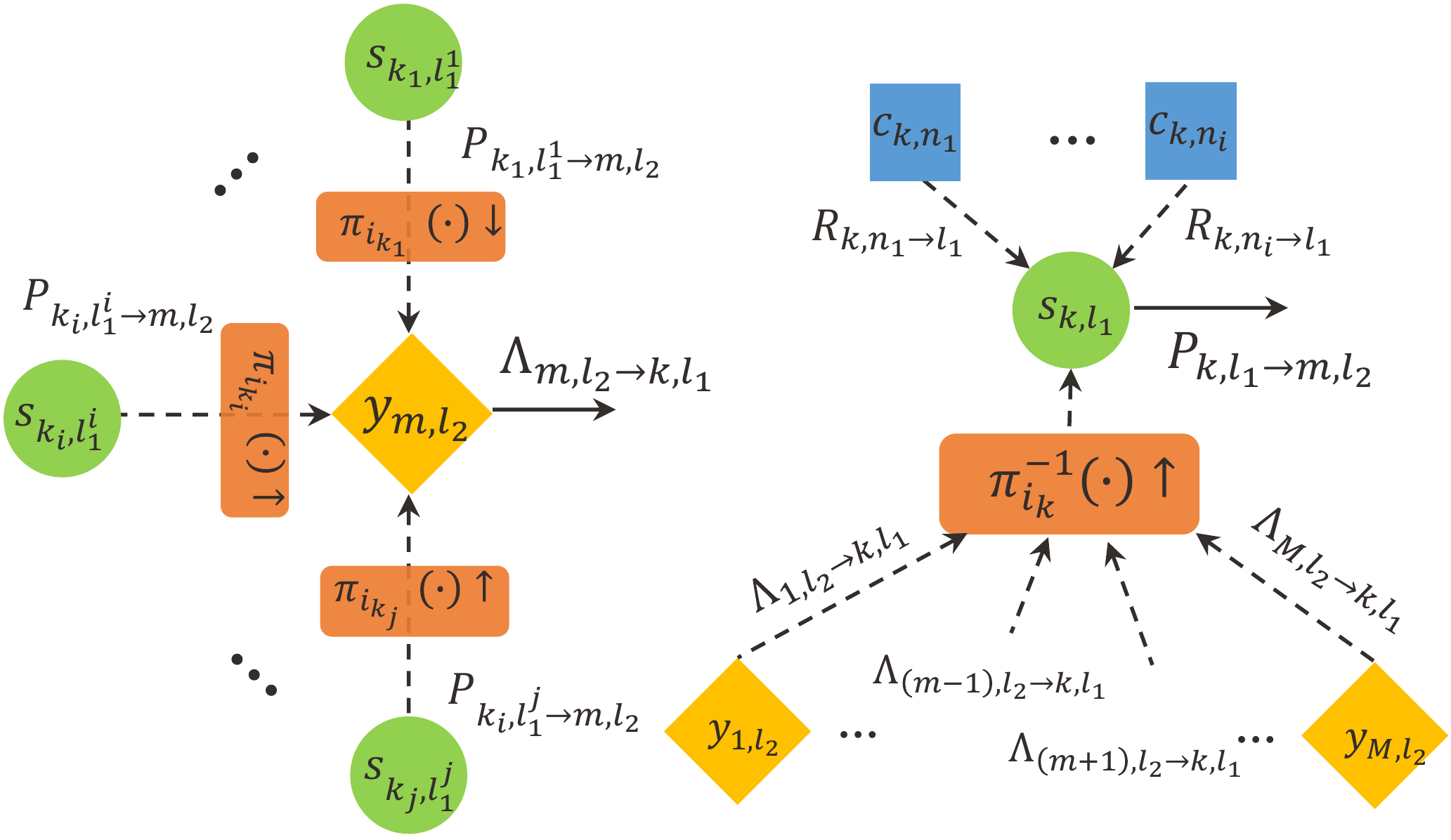}}
	\caption{Update rules for messages $\Lambda$ and $P$ at VNs and SNs on the factor graph. The dashed and solid lines represent input and output messages, respectively.}
	\label{pic-6}
\end{figure}

\par  The message ${\Lambda _{m,l_2 \to k,l_1}}$ in \eqref{equ-36} is the LLR with the probability of VN $s_{k,l_1}$ taking different values observed at SN $y_{m,l_2}$. For BPSK modulated system, it is given by
\begin{equation}
	\begin{aligned}
	{\Lambda _{m,l_2 \to k,l_1}} &= \log \frac{{P\left( {{y_{m,l_2}}|\mathbf{H},{s_{k,l_1}} = +1} \right)}}{{P\left( {{y_{m,l_2}}|\mathbf{H},{s_{k,l_1}} =  - 1} \right)}} \\
&= \frac{2}{{\sigma _{{z_{k,l_2}}}^2}}\text{Re}\left( {h_{m,k}^ * \left( {{y_{m,l_2}} - {\mu _{{z_{k,l_2}}}}} \right)} \right),
\label{equ-38}
	\end{aligned}
\end{equation}
where $\bm{H}$ is the channel matrix, which can be estimated in the CS phase. We note that similar to the joint DAD-CE algorithm, \eqref{equ-38} is also obtained by treating the interference from other devices as noise. We rewrite \eqref{equ-35} as
\begin{equation}
	\begin{aligned}
			{y_{m,l_2}} &= \sum\limits_{\substack{j \in \mathcal{K}(m,l_2),l \in \mathcal{L}(m,l_2)}} {{h_{m,j}}{\pi_{i_j}\left( \tilde{s}_{j,l}\right) } + {n_{m,l_2}}}\\
		&={h_{m,k}}\pi_{i_k}\left( \tilde{s}_{k,l_1}\right)  \!+\! \underbrace{\sum\limits_{\substack{j \in \mathcal{K}(m,l_2) \backslash  k\\l \in \mathcal{L}(m,l_2)\backslash l_1}} {{h_{m,j}}\pi_{i_j}\left( \tilde{s}_{j,l}\right)  \!+\! {n_{m,l_2}}}}_{z_{k,l_2}}, \label{equ-40}
	\end{aligned}
\end{equation}
where $n_{m,l_2}$ is the Gaussian noise with zero mean and variance ${\sigma _n^2}$, and $\mathcal{K}(m,l_2),\mathcal{L}(m,l_2)$ denote the set of devices as well as the corresponding bits related with SN $y_{m,l_2}$, respectively. Accordingly, the set of related VNs is $\left\lbrace s_{j,l} \vert j \in \mathcal{K}(m,l_2), l \in \mathcal{L}(m,l_2) \right\rbrace $ and we note that the subscripts $j$ and $l$ are one-to-one mappings. For instance, Fig. \ref{pic-7} showcases the VNs related with SN $y_{1,3}$. In this regard, $\mathcal{K}(1,3) = \left\lbrace 1,3,5 \right\rbrace $ and $\mathcal{L}(1,3)= \left\lbrace5,4,2 \right\rbrace $ and the corresponding VNs are $\left\lbrace  s_{1,5}, s_{3,4}, s_{5,2} \right\rbrace $. The Gaussian noise $n_{m,l_2}$ and the interference from other devices are all treated as noise denoted by $z_{k,l_2}$, which is a Gaussian variable with mean ${\mu _{{z_{k,l_2}}}}$ and variance ${\sigma _{{z_{k,l_2}}}^2}$ given by
\begin{align}
	{\mu _{{z_{k,l_2}}}}    &= \sum\limits_{\substack{j \in \mathcal{K}(m,l_2) \backslash  k\\l \in \mathcal{L}(m,l_2)\backslash l_1}} {{h_{m,j}} \cdot \mathbb{E}\left[ \pi_{i_j}\left( \tilde{s}_{j,l}\right)  \right] }  \label{equ-41} \\ 
	\sigma _{{z_{k,l_2}}}^2 &= \sum\limits_{\substack{j \in \mathcal{K}(m,l_2) \backslash  k\\l \in \mathcal{L}(m,l_2)\backslash l_1}} {{{\left| {{h_{m,j}}} \right|}^2}\cdot  \text{Var}\left[ \left(\pi_{i_j}\left( \tilde{s}_{j,l}\right)  \right)\right] }+\sigma_n^2. \label{equ-42}	
\end{align}
And for BPSK modulation, the mean and variance of $\pi_{i_j}\left( \tilde{s}_{j,l}\right) $ is given by
\begin{align}
	\mathbb{E}\left[ \pi_{i_j}\left( \tilde{s}_{j,l}\right)  \right]  &=2 \cdot \pi_{i_j}\left({P_{j,l \to m,l_2}}\right)  - 1 \\
	\text{Var}\left[ \left(\pi_{i_j}\left( \tilde{s}_{j,l}\right)  \right)\right]  &= 4\cdot\pi_{i_j}\left( \left( {1 - {P_{j,l \to m,l_2}}} \right) \cdot {P_{j,l \to m,l_2}}\right),
\end{align}
where ${P_{j,l \to m,l_2}}$ denotes the probability of VN $s_{j,l}=1$, and is initialized to $0.5$. We note that it needs to be interleaved before the calculation. As shown in Fig. \ref{pic-6}, ${P_{k,l_1 \to m,l_2}}$ is updated by collecting the incoming messages from CNs related with VN $s_{k,l_1}$ and those from all SNs except $y_{m,l_2}$.  We give the update rule as below.
\begin{equation}
	{P_{k,l_1 \to m,l_2}} \!=\! \frac{{\exp \!\left( \Lambda + R\right)}}{{1 + \exp \left( \Lambda + R\right)}}, \label{equ-43}
\end{equation}
where
\begin{equation}
	\Lambda =\!\!\! \sum\limits_{j \in M \backslash m} {\pi_{i_j}^{-1}\left({\Lambda _{j,l_2 \to k,l_1}}\right) }, ~~R=\!\!\!\sum\limits_{j \in {\mathcal{N}_c}\left( {k,l_1} \right)} {{R_{k,j \to l_1}}}. \label{equ-add-1}
\end{equation}
Similar to \eqref{equ-36}, the message $\Lambda _{j,l_2 \to k,l_1}$ needs to be de-interleaved in the update of $P$. The LLR of VN $s_{k,l_1}$ at the end of an iteration is given by
\begin{equation}
	{L_{k,l_1}} = \sum\limits_{j \in M} {\pi_{i_j}^{-1}\left( {\Lambda _{j,l_2 \to k,l_1}}\right)  + \sum\limits_{j \in {\mathcal{N}_c}\left( {k,l_1} \right)} {{R_{k,j \to l_1}}} }.  \label{equ-46}
\end{equation}

\par The information bit $\hat{v}_{k,l_1}^c$ is decoded as one if $L_{k,l_1}>0$ and zero otherwise. Since LDPC codes are described by the parity matrix $\bm{C}$, the iteratively decoding process is continued till $\text{mod} \left( \bm{C}\hat{\bm{v}}_k^c,2\right) =0$ or the maximum number of iterations $N_{iter}$ is reached. 

\par To further improve the spectrum efficiency, the QPSK modulation is also considered in this coding system, of which the constellation set is $\mathbb{S}=\left\lbrace \pm 1/  \sqrt{2},\pm 1/ \sqrt{2} i\right\rbrace $. Briefly, QPSK modulated signals can be split into two orthogonal BPSK ones. As such, we can implement the above MP algorithm on these two signals separately. Additionally, the real and imaginary parts of the messages $\Lambda$ and $P$ need to be considered separately, so does the parity of subscript $l_1$ in $Q_{k,l_1 \to n}$ and $R_{k,n \to l_1}$. It is worth noting that the range of $l_1$ in $\Lambda$ and $P$ is half of that in $Q$ and $P$, i.e., $\left[1:\tilde{L}_c/2 \right]$, since there are both messages on the real and imaginary parts. In what follows we give the updated rules of these messages.

\par Similar to \eqref{equ-38}, $\Lambda$ is the LLR of the probability that $s_{k,l_1}$ takes different values in $\mathbb{S}$. However, it is no longer a real number. Instead, it is given by
\begin{equation}
\Lambda _{m,l_2 \to k,l_1}	= \frac{2\sqrt{2}}{{\sigma _{{z_{k,l_2}}}^2}}\cdot  {h_{m,k}^ *\cdot  \left( {{y_{m,l_2}} - {\mu _{{z_{k,l_2}}}}} \right)},  \label{equ-47}
\end{equation}
where the mean and variance of $z_{k,l_2}$ are given in \eqref{equ-41} and \eqref{equ-42}, respectively. And for QPSK modulation, the mean and variance of $\pi_{i_j}\left( \tilde{s}_{j,l}\right) $ is given by
\begin{align}
	\mathbb{E}\left[ \pi_{i_j}\left( \tilde{s}_{j,l}\right)  \right]  &= 1/\sqrt{2} \cdot \left\lbrace  2\cdot \pi_{i_j} \left(  {P_{j,l \to m,l_2}^r}\right)    - 1 \right.  \notag \\ 
	  & \quad \quad \quad ~\left. +\left[ 2\cdot \pi_{i_j} \left(  {P_{j,l \to m,l_2}^i}\right)  -1 \right] \cdot i\right\rbrace   \label{equ-48}\\ 
	\text{Var}\left[ \left(\pi_{i_j}\left( \tilde{s}_{j,l}\right)  \right)\right]  &= 2\cdot\pi_{i_j}\left[ {P_{j,l \to m,l_2}^r} - \left({P_{j,l \to m,l_2}^r} \right)^2 \right. \notag \\
	&\quad \quad \quad ~ \left. + {P_{j,l \to m,l_2}^i} - \left({P_{j,l \to m,l_2}^i} \right)^2 \right] , \label{equ-49}
\end{align}
where $ {P_{j,l \to m,l_2}^r}$ and $ {P_{j,l \to m,l_2}^i}$ are the real and imaginary parts of $P_{j,l \to m,l_2}$, respectively. Similar to \eqref{equ-43} and \eqref{equ-add-1}, $ {P_{k,l_1\to m,l_2}^r}$ and $ {P_{k,l_1\to m,l_2}^i}$ are given by
\begin{align}
		{P_{k,\lceil \frac{l_1}{2} \rceil \to m,l_2}^r} &=  \frac{\exp \left[  \text{Re}\left( \Lambda \right) + R  \right]}{{1 + \exp \left[\text{Re}\left( \Lambda \right) + R \right]}}, ~l_1~\text{is odd}, \label{equ-51} \\
		{P_{k,\frac{l_1}{2} \to m,l_2}^i} &=  \frac{\exp \left[  \text{Im}\left( \Lambda \right) +R \right]}{{1 + \exp \left[\text{Im}\left( \Lambda \right) +R  \right]}}, ~l_1~\text{is even}. \label{equ-52} 
\end{align}
As such, the range of subscript $l_1$ in $P$ is half of that in $R$ as aforementioned. $\Lambda$ and $R$ are defined in \eqref{equ-add-1}. The update rule for message $R$ is the same as \eqref{equ-37} and that for $Q$ is given by 
\begin{equation}
	{Q_{k,l_1 \to n}} = \left\{\begin{array}{l}
		\sum\limits_{j \in M} {\pi_{i_j}^{-1}\left(\text{Re}\left(  \Lambda _{j,l_2 \to k,\lceil \frac{l_1}{2} \rceil}\right) \right) } \! \\
		\quad \quad \quad  + \!\!  \sum\limits_{j \in {\mathcal{N}_c}\left( {k,l_1} \right) \backslash n} {\!\!\!{R_{k,j \to l_1}}}, ~l_1~\text{is odd}\\
		\sum\limits_{j \in M} {\pi_{i_j}^{-1}\left(\text{Im}\left(  \Lambda _{j,l_2 \to k,\frac{l_1}{2}}\right) \right) } \! \\
		\quad \quad \quad + \!\!  \sum\limits_{j \in {\mathcal{N}_c}\left( {k,l_1} \right) \backslash n} {\!\!\!{R_{k,j \to l_1}}}, ~l_1~\text{is even}\\
	\end{array} \right.  \label{equ-53}
\end{equation}
The LLR of VN $s_{k,l_1}$ at the end of an iteration is given by
\begin{equation}
	L_{k,l_1} = {Q_{k,l_1 \to n}} + R_{k,n \to l_1}, ~ \forall n\in\mathcal{N}_c(k,l_1). \label{equ-54}
\end{equation}
The decision rule and termination condition are the same as those in the BPSK system mentioned earlier. Finally, we obtain the decoded messages.

\par Note that the estimated number of active devices $K$ is not guaranteed to be equal to $K_a$. Therefore, not all the decoded messages satisfy the parity check. We denote $\widehat{\mathcal{V}}=\left\{\hat{\bm{v}}_k^c, k \in \widehat{\mathcal{K}}\right\}$ and  $\widehat{\mathcal{K}}$ as the set of successfully decoded messages and the  corresponding devices, respectively. And we have $\left| {\widehat {\mathcal{K}}} \right| \le {K_a}$. To further improve the performance, we combine the MIMO-LDPC decoder with the SIC method and we denote it as the MIMO-LDPC-SIC algorithm, which works as follows.

\par Let $\widehat {\mathbf{H}} \in {C^{K \times M}}$ and $\mathcal{K}$ denote the channel matrix and the set of active devices estimated in the CS phase. Let ${\hat {\mathcal{V}}_0}$ and ${\hat {\mathcal{K}}_0}$ respectively denote the sets of correctly decoded messages (i.e., those that satisfy the check) and the corresponding devices obtained by the decoder, which are initialized to empty sets. With $\bm{Y}_c$, interleaving patterns $\left\lbrace \pi_{i_k}, k \in \mathcal{K}\right\rbrace $ and $\hat{\bm{H}}$, the decoder outputs the set of successfully decoded messages  ${\widehat {\mathcal{V}}} $ and the corresponding devices ${\widehat {\mathcal{K}}}$. Then we have ${\widehat {\mathcal{V}}_0} \leftarrow {\widehat {\mathcal{V}}_0} \cup \widehat {\mathcal{V}}$, ${\widehat {\mathcal{K}}_0} \leftarrow {\widehat {\mathcal{K}}_0} \cup \widehat {\mathcal{K}}$ and ${\mathbf{H}} = {\widehat {\mathbf{H}}_{k,:}}$ for $k \in {\mathcal{K}} \backslash {\widehat {\mathcal{K}}_0}$. The residual signal is updated by
\begin{equation}
	\mathbf{Y} = {\mathbf{Y}_c} - \sum\nolimits_{k \in {{\hat {\mathcal{K}}}_0}} {{\pi _{{i_k}}}\left( {{\tilde{\mathbf{s}}_k}} \right){\widehat{\mathbf{H}}_{k,:}} },
	\label{equ-55}
\end{equation}
where $\tilde{\mathbf{s}}_k \in \mathbb{C}^{L_c\times 1}$ is the $k$-th codeword in $\widehat {\mathcal{V}}_0$ after modulation and zero-padding. The updated $Y$ and $\bm{H}$ as well as the interleaving patterns are sent to the MIMO-LDPC decoder for the next round of decoding. This iterative process ends when ${\hat {\mathcal{K}}} = \emptyset $ or ${\mathcal{K}} \backslash {\widehat {\mathcal{K}}_0} =\emptyset$. The overall decoding algorithm is summarized in Algorithm \ref{alo-3}.

  \begin{algorithm}  
	\caption{MIMO-LDPC-SIC Decoding Algorithm}
	\label{alo-3}  
	\begin{algorithmic}[1]
		\STATE {{\bf Input}: $\bm{Y_c}$, $\hat{\bm{H}}$, $\mathcal{L}= \left\{ {{i_k},k \in \mathcal{K}} \right\}$, $\sigma_n^2$}\\
		\STATE {{\bf Initialize}:\\ $\bm{Y} \!=\!\bm{Y_c}$, $\bm{H}\!=\!\hat{\bm{H}}$, ${\widehat {\mathcal{V}}_0} \!\leftarrow\! \emptyset $, ${\widehat {\mathcal{K}}_0} \!\leftarrow\! \emptyset $, ${R_{k,n \to l_1}}=0$,\\ BPSK: ${P_{k,l_1 \to m,l_2}}=0.5$, QPSK: ${P_{k,l_1 \to m,l_2}}=0.5 + 0.5i$.}\\
		\REPEAT
		\REPEAT
			\STATE{$\Lambda$ update: $\Lambda _{m,l_2 \to k,l_1}$ by \eqref{equ-38} or \eqref{equ-47}.}\\
			\STATE{$Q$ update: $Q_{k,l_1 \to n}$ by \eqref{equ-36} or \eqref{equ-53}.}\\
			\STATE{$R$ update: $R_{k,n \to l_1}$ by \eqref{equ-37}.}\\
			\STATE{$P$ update: $P_{k,l_1 \to m,l_2}$ by \eqref{equ-43} or \eqref{equ-51}-\eqref{equ-52}.}\\
			\STATE{$L$ update and hard decision: $	L_{k,l_1}$ by \eqref{equ-46} or \eqref{equ-54}.}\\
		\UNTIL{$N_{iter}$ reached or $\text{mod}(\bm{C}\hat{\bm{v}}^c,2)=0$.}
		\STATE{{\bf Output}: ${\widehat {\mathcal{V}}} $,  ${\widehat {\mathcal{K}}}$}\\
		\STATE{${\widehat {\mathcal{V}}_0} \leftarrow {\widehat {\mathcal{V}}_0} \cup \widehat {\mathcal{V}}$, ${\widehat {\mathcal{K}}_0} \leftarrow {\widehat {\mathcal{K}}_0} \cup \widehat {\mathcal{K}}$.}\\
		\STATE{${\mathbf{H}} = {\widehat {\mathbf{H}}_{k,:}}$ for $k \in {\mathcal{K}} \backslash {\widehat {\mathcal{K}}_0}$.}\\
		\STATE{$	\mathbf{Y} = {\mathbf{Y}_c} - \sum\nolimits_{k \in {{\hat {\mathcal{K}}}_0}} {{\pi _{{i_k}}}\left( {{\tilde{\mathbf{s}}_k}} \right){\widehat{\mathbf{H}}_{k,:}} }$.}\\
		\UNTIL{${\widehat {\mathcal{K}}} = \emptyset $ or ${\mathcal{K}} \backslash {\widehat {\mathcal{K}}_0} =\emptyset$.}\\
		\STATE{{\bf Return:} ${\widehat {\mathcal{V}}_0}$, ${\widehat {\mathcal{K}}_0}$}	
	\end{algorithmic}  
\end{algorithm} 

\par We note that the stitching of the messages in the CS and LDPC phases is easy. In URA, devices are not identified, as such, the IDs can not be employed to distinguish the messages. We have known that the interleaving patterns and channels $\left\lbrace \pi_{i_k}, \hat{\bm{h}}_k, k \in \widehat{\mathcal{K}}_0\right\rbrace $ obtained in the CS phase acting as key parameters participate in the decoding process in the LDPC phase. For each decoded message in  $\widehat{\mathcal{V}}_0$, it is decoded with a specific interleaving pattern $\pi_{i_k}$ as well as the channel $\hat{\bm{h}}_k$. As aforementioned,  $\pi_{i_k}$ is uniquely determined by the message index representation $i_k$, which directly corresponds to device $k$' preamble. Therefore, $\pi_{i_k}$ and $\hat{\bm{h}}_k$ establish a connection of devices' messages in the two phases. Briefly, if $\hat{\bm{v}}^c$ is successfully decoded with the participation of $\pi_{i_k}$ and $\hat{\bm{h}}_k$, it is exactly the latter $B_c$ bits of message of device $k$. As such, the stitching of the messages in two phases will not be a problem.
 
 \subsection{Joint Update} \label{sec-5-3}
 \par The above CS and LDPC decoders can recover the  $B$ bits of information with their work in tandem. Besides, thanks to the consistency of the above MP algorithm, the BP-based decoder can draw a connection of the decoding process in the CS and LDPC phases. That is, messages in the CE as well as the MIMO-LDPC decoding processes can be jointly updated by utilizing the belief of each other, thus leading to improved performance. This joint update algorithm is denoted as joint DAD-CE-DD algorithm and elaborated as follows. 
 
 \par For the successfully decoded devices ${\widehat {\mathcal{K}}_0}$, the corresponding messages $\left\lbrace {\pi _{{i_k}}}\left( {{\tilde{\mathbf{s}}_k}} \right), k \in \widehat {\mathcal{K}}_0 \right\rbrace $ can be leveraged as soft pilot sequences joint with their codewords $\left\lbrace \bm{a}_{i_k},  k \in \widehat {\mathcal{K}}_0 \right\rbrace $ in the CS phase to carry out a second CE. This longer pilot sequence will lead to a better CE performance, which has been confirmed in  our simulation in Fig. \ref{pic-11}. We note that the subsequent CE is conducted via the above joint DAD-CE algorithm with devices' activity fixed. In this regard, the Gaussian messages $\left\lbrace \mu_{lm \to km}^{SN}, \bm{\Sigma}_{l \to k}^{SN}, \mu_{lm \to km}^{VN}, \bm{\Sigma}_{l \to k}^{VN}\right\rbrace $ in \eqref{equ-18}-\eqref{equ-19} and \eqref{equ-26}-\eqref{equ-27} are iteratively updated with a longer observation sequence $\left\lbrace \bm{y}_l, l\in \left[ 1:L_p+L_c\right] \right\rbrace$. Besides, it is worth noting that in the CE output in \eqref{equ-30}-\eqref{equ-31}, the prior mean and covariance $\mu_{k}^{pri}, \bm{\Sigma}_{k}^{pri}$ are no longer zero and $\bm{I}_M$, respectively. Instead, they are the output estimation $\mu_{k}^{dec}$ and estimating deviation $\bm{\Sigma}_{k}^{dec}$ of $\bm{h}_k$ in the first CE, respectively.
 
   \begin{algorithm}  
 	\caption{Joint DAD-CE-DD Algorithm}
 	\label{alo-4}  
 	\begin{algorithmic}[1]
 		\STATE {{\bf Input}: $\bm{Y_p}$, $\bm{Y_c}$, $\bm{A}$, $\bm{\mu}^{pri}$, $\bm{\Sigma}^{pri}$, $\sigma_n^2$, $p_a$}\\
 		\STATE {$ \triangleright $ CS Phase: \\\quad  Joint DAD-CE Algorithm}\\
 		\STATE{\quad $ \triangleright $ Collision Resolution Protocol}\\
 		\STATE{\quad Output: $\bm{\mu}^{dec}$, $\bm{\Sigma}^{dec}$, $\mathcal{L}= \left\{ {{i_k},k \in \mathcal{K}} \right\}$}\\
 		\STATE {$ \triangleright $ Joint DAD-CE-DD Algorithm:}\\
 		\STATE{\quad Initialize: $\widetilde{\bm{Y}}_r= \bm{Y}_c$, $\widetilde{\mathcal{V}}=\emptyset$}, $\widetilde{\mathcal{K}}=\emptyset$\\		
 		\STATE {\quad \bf repeat}
 		\STATE {\quad \quad  $ \triangleright $ LDPC Phase:} 
 		\STATE{\quad \quad \quad  Input: $\widetilde{\bm{Y}}_r$, $\bm{\mu}^{dec}$, $\mathcal{L}$, $\sigma_n^2$}\\
 		\STATE{\quad \quad \quad MIMO-LDPC-SIC Decoding Algorithm}\\
 		\STATE{\quad \quad \quad Output: ${\widehat {\mathcal{V}}_0}$, ${\widehat {\mathcal{K}}_0}$}\\
 		\STATE{\quad \quad{${\widetilde {\mathcal{V}}} \leftarrow {\widetilde {\mathcal{V}}} \cup \widehat {\mathcal{V}}_0$, ${\widetilde {\mathcal{K}}} \leftarrow {\widetilde {\mathcal{K}}} \cup \widehat {\mathcal{K}}_0$}}
 		\STATE{\quad \quad $\widetilde{\bm{Y}}_c =  \sum\nolimits_{k \in {{\widehat {\mathcal{K}}}_0}} {{\pi _{{i_k}}}\left( {{\tilde{\mathbf{s}}_k}} \right){\bm{\mu}^{dec}_k }}+ \bm{Z}$}\\
 		\STATE{ \quad \quad $ \triangleright $ CS Phase:}\\
 		\STATE{\quad \quad \quad  Input: $\bm{Y}_p$, $\bm{A}$, $\widetilde{\bm{Y}}_c$, ${\widehat {\mathcal{V}}_0}$, $\bm{\mu}^{dec}$, $\bm{\Sigma}^{dec}$, $\sigma_n^2$}\\
 		\STATE{\quad \quad \quad CE (Activity fixed)}\\
 		\STATE{\quad \quad \quad $ \triangleright $ Collision Resolution Protocol }\\
 		\STATE{\quad \quad \quad Output: $\widetilde{\bm{\mu}}^{dec}$, $\widetilde{\bm{\Sigma}}^{dec}$ }\\
 		\STATE{\quad \quad  $\widetilde{\bm{Y}}_r = {\mathbf{Y}_c} - \sum\nolimits_{k \in {{\widetilde {\mathcal{K}}}}} {{\pi _{{i_k}}}\left( {{\tilde{\mathbf{s}}_k}} \right){\widetilde{\bm{\mu}}^{dec}_k} }$}
 		\STATE {{\quad  \bf until} ~ ${\widehat {\mathcal{K}}}_0 = \emptyset $ or ${\mathcal{K}} \backslash {\widetilde {\mathcal{K}}} =\emptyset$}\\
 		\STATE{{\bf Return:} $\widetilde{\mathcal{V}}$, $\widetilde{\mathcal{K}}$, $\mathcal{L}= \left\{ {{i_k},k \in \widetilde{\mathcal{K}}} \right\}$, $\widetilde{\bm{\mu}}^{dec}$}	
 	\end{algorithmic}  
 \end{algorithm} 
 
\emph{Remark 1:} We remark that messages in the MP algorithm exhibit the property of consistency and unity. As such, messages among different parts can always be jointly processed and updated. For instance, recalling the Joint DAD-CE algorithm, where the Bernoulli messages are updated by utilizing the Gaussian messages. Likewise, in the MIMO-LDPC-SIC Decoding algorithm, the MP algorithm can be applied to MIMO detection or LDPC decoding. In the proposed algorithm, we combine these two parts and update the underlying messages jointly, i.e., the LLR message of the symbol in MIMO detection can be involved in the LDPC decoding process, and vice versa. Moreover, the property is again exploited in the Joint DAD-CE-DD algorithm. By leveraging the belief of the estimated channel and the correctness of the LDPC codewords, the CE can be again performed aided with the correctly decoded codewords in the LDPC phase, thus connecting these two phases. Throughout the paper, we take into consideration the idea of the joint update for the messages in MP-based algorithms.
 
 \par
\begin{figure}[htpb]
 	\centerline{\includegraphics[width=0.5\textwidth]{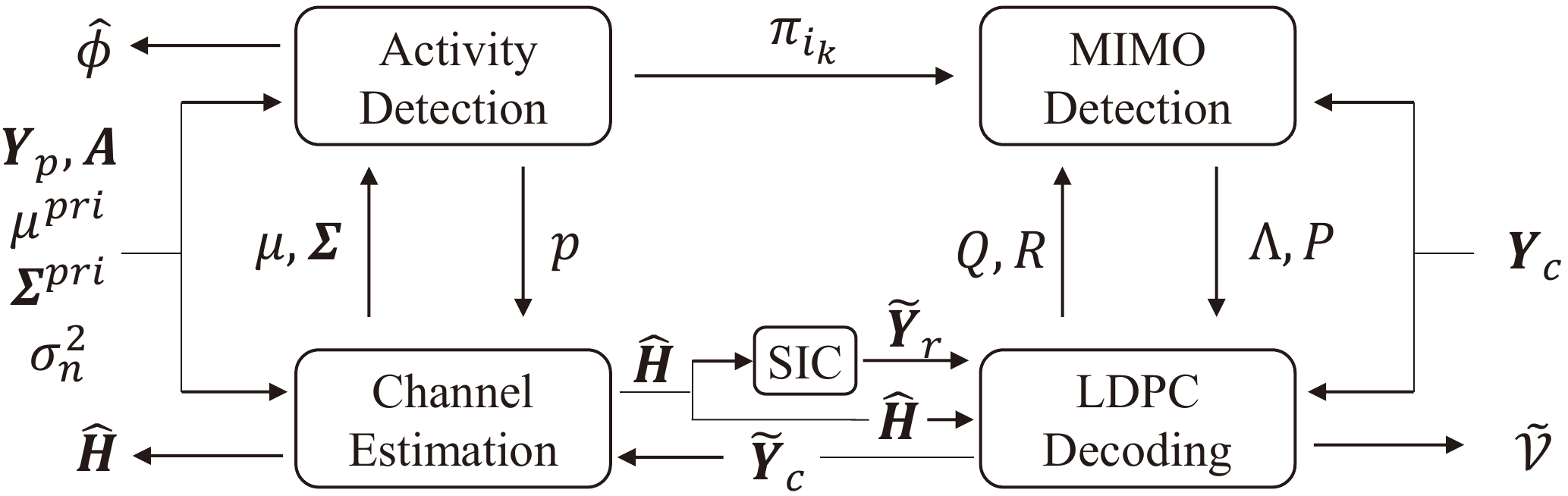}}
 	\caption{The diagram of the joint DAD-CE-DD algorithm.}
 	\label{pic-8}
 \end{figure}

 One might argue that for the correctly decoded messages, improving the accuracy of the corresponding channels will not bring substantial performance improvements. Nevertheless, according to the SIC method in \eqref{equ-55}, by improving the channel accuracy of those devices whose messages are successfully decoded, the residual signal of the incorrectly decoded messages can be obtained more accurately. We denote the residual signal with improved accuracy as $\widetilde{\bm{Y}}_r$. Consequently, $\widetilde{\bm{Y}}_r$ will improve the performance for those messages that have not been successfully decoded yet. And this is the reason why the channel needs to be estimated twice or more times. We give a brief diagram for this algorithm in Fig. \ref{pic-8}, where $\widetilde{\bm{Y}}_c$ denotes the noisy signal reconstructed from the correctly decoded messages.
 
\par We note this data-aided CE algorithm appears to be similar to the work in \cite{hiroki2020arxiv}, which employs the data as soft pilots to conduct CE jointly with the real pilot sequences. However, the convergence and correctness of the estimated data cannot be verified, which may result in the propagation of errors and failure of the joint data and channel estimation. Whereas the convergence of the proposed algorithm can be guaranteed. The proposed Joint DAD-CE-DD algorithm consists of two modules, namely, the Joint DAD-CE algorithm for activity detection and channel estimation, and the MIMO-LDPC-SIC Decoding algorithm for data decoding. Messages are passed between these two modules. For arbitrary channel estimation results as input, the decoding process will be executed by the MIMO-LDPC-SIC Decoding algorithm. Once a codeword is successfully decoded (i.e., satisfies the check), it will not be changed in subsequent iterations. Therefore, the number of correctly decoded codewords is monotonically non-decreasing at each iteration. Correspondingly, the error rate $P_e$ is monotonically non-increasing, and also, it is bounded below by zero. According to the well-known monotone convergence theorem \cite{converge}, the proposed algorithm is guaranteed to converge. Besides, with the correctly decoded codewords aided to estimate the channel, an improved CE accuracy can be guaranteed and it contributes to a higher probability of successfully decoding for those who have not yet. This iterative algorithm ends when no more messages are correctly decoded or all the messages are decoded successfully, which is summarized in Algorithm \ref{alo-4}.

\subsection{Complexity}
 \par In this section, we compare the complexity of the proposed algorithm with the existing alternative in terms of the real-number multiplication (or division), addition (or subtraction), and other operations (eg., exp, log, tan, $\text{tan}^{-1}$), as shown in Table \ref{tab-2}. As a divide-and-conquer strategy, there are inner and outer decoders in the CB-ML algorithm, referred to as ML and tree decoder, respectively. The complexity of our algorithm is mainly incurred by the Joint DAD-CE algorithm and MIMO-LDPC decoder. In Table \ref{tab-2}, the parameters $\alpha$ and $c$ satisfy that $B_p=\alpha \log_2K_a, L_p=c\log_2K_a$, respectively. We note that the complexity of the tree decoder is defined by the total number of parity check constraints that need to be verified \cite{Amalladinne2020tit}, and this is obtained in the regime that $B_p$ and $K_a$ tend to infinity. Nevertheless, in practice, it increases exponentially with the number of slots $S$. Besides, the tree code in \cite{Fengler2021tit} verifies the estimated codewords in each segment successively, and the check and estimate are performed separately. While during the LDPC decoding process in our algorithm, the check and estimate of the codeword are performed jointly and simultaneously, both by calculating the soft messages. Once the codeword is estimated, the corresponding check is completed. As for the inner code, the complexity of the ML decoder is nearly the same order as that of our algorithm. However, as a consequence of the coordinate descent algorithm, there are $2^{B_p}$ cycles in the ML decoding per iteration, which can only be computed successively. On the contrary, as a remarkable property of the MP algorithm, all computations in our algorithm can be decomposed into many local ones, which can be performed in parallel over the factor graph. Hence, our algorithm has lower time complexity. In general, the proposed algorithm exhibits a complexity linear with $K, M, L$, which is lower than the existing CB-ML algorithm.

 \begin{table*}[!t]
	\caption{COMPLEXITY COMPARISON PER ITERATION }
		\renewcommand{\arraystretch}{1.7}
		\setlength{\tabcolsep}{4.2mm}{
		\begin{tabular}{l|l|l|l|l}
			\hline
			\multicolumn{2}{l|}{\bf Algorithms} & {\bf Real Multi. / Div.} &{ \bf Real Add. / Sub. }& {\bf Others } \\
			\hline
			\multirow{2}{*}{\tabincell{c}{CB-ML}} & ML decoder& $\mathcal{O}(KL_p^2)$ & $\mathcal{O}(KL_p^2)$ & $\backslash$\\
			\cline{2-5}
			& Tree decoder & \multicolumn{2}{l|}{$\mathcal{O}\left(K_a^{\alpha/c}\text{log}_2K_a\right), B_p, K_a \rightarrow \infty $} & $\backslash$ \\
			\hline
			\multirow{2}{*}{\tabincell{c}{Joint DAD-CE-DD}} &Joint DAD-CE & $\mathcal{O}(KML_p)$ & $\mathcal{O}(KML_p)$ & exp / log: $\mathcal{O}(KL_p)$\\
			\cline{2-5}
			& MIMO-LDPC& $\mathcal{O}( \widehat{K}M\widetilde{L}_c) $ & $\mathcal{O}( \widehat{K}M\widetilde{L}_c) $ & exp: $\mathcal{O}( \widehat{K}M\widetilde{L}_c) $,  tanh / $\text{tanh}^{-1}$: $\mathcal{O}(\widehat{K}NN_v)$  \\
			\hline	
			\multicolumn{5}{l}{\tabincell{l}{Note: $K=2^{B_p}$, $\widehat{K}, N, N_v$ denote the estimated number of active devices, the rows and row weights of the parity check matrix, respectively.}}	
	\end{tabular}}
	\label{tab-2}
\end{table*}

 \section{Numerical Results} \label{simulation}
 \subsection{Parameter Settings}
 \par In this section, we assess the overall performance of the proposed framework with the metric defined in \eqref{equ-3}-\eqref{equ-4}. The CB-ML proposed by Fengler et. al. in \cite{Fengler2021tit} serves as the baseline in this paper. For the sake of fair comparison to the benchmarks, and isolating the fundamental aspects of the problem without additional model complication, we consider the flat path loss model in the simulation, i.e., the channel is i.i.d. Rayleigh fading model and the LSFC is fixed to $\beta_k=1$ in all schemes. We note again that this can be achieved by the well-studied power control schemes \cite{Chandrasekhar2009twc,Patel1994jsac,Turin1984jsac} in practice. As such, the carrier frequency is not specified in the simulation, since it can be arbitrary and does not affect the performance of the proposed algorithm, when considering flat path loss model. The user distribution is considered to be uniformly distributed in the cell. The noise variance is set to $\sigma_n^2=1$ and is known to all schemes. The maximum scheduling times for the collision resolution protocol is $t_{max}=3$. For the joint DAD-CE algorithm, MIMO-LDPC-SIC decoder, and Joint DAD-CE-DD algorithm, the maximum number of iterations is set to $N_{iter}=20$, $30$, and $20$, respectively. We fix the message length to $B=96$ with $B_p=12$ and $B_c=84$. The length of the CS codeword is $L_p=100$ in both schemes. In the channel coding part, we employ the $(3,6)$-regular LDPC code \cite{LDPC} with the rate $0.5$. In our scheme, the length of channel use varies with the number of zeros padded in the sequence. However, it is discretely valued in the CB-ML scheme and changes with the number of slots denoted as $S$. Since the length of the information is fixed, we can change $S$ by adjusting the parity check allocation, which is given in Table \ref{tab-1} according to the principle in \cite{Amalladinne2020tit}. We note that the slot length $J=12$ aligns with $B_p$.
  
 \begin{table}[h]
 	\centering
 	\caption{The parity check allocation for different number of slots.}
 	\setlength{\tabcolsep}{8mm}{
 		\begin{tabular}{l l}
 			\toprule
 			$S$ & Parity Check Allocation \\
 			\midrule
 			12 & (12,0), (3,9), (9,3), $\cdots$, (9,3), (0,12)\\
 			\specialrule{0em}{0.5pt}{0.5pt} 
 			13 & (12,0), (4,8), (8,4), $\cdots$, (0,12)\\
 			\specialrule{0em}{0.5pt}{0.5pt} 
 			14 & (12,0), (7,5), (7,5), $\cdots$, (7,5), (0,12)\\
 			\specialrule{0em}{0.5pt}{0.5pt} 
 			15 & (12,0), (7,5), (7,5), $\cdots$, (7,5), (0,12), (0,12)\\
 			\specialrule{0em}{0.5pt}{0.5pt} 
 			16 & (12,0), (6,6), (6,6), $\cdots$, (6,6), (0,12)\\
 			\specialrule{0em}{0.5pt}{0.5pt} 
 			17 & (12,0), (6,6), (6,6), $\cdots$, (6,6), (0,12), (0,12)\\			
 			\bottomrule
 	\end{tabular}}
 	\label{tab-1}
 \end{table}
 
 \par In Table \ref{tab-1}, (3,9) means that the first three are data bits and the last nine are parity bits and so on. For more details, we refer the reader to \cite{Fengler2021tit} for CB-ML algorithm and \cite{Amalladinne2020tit} for the tree coding scheme.
 
  \subsection{Results}
  
    \begin{figure}[htpb]
  	\centerline{\includegraphics[width=0.46\textwidth]{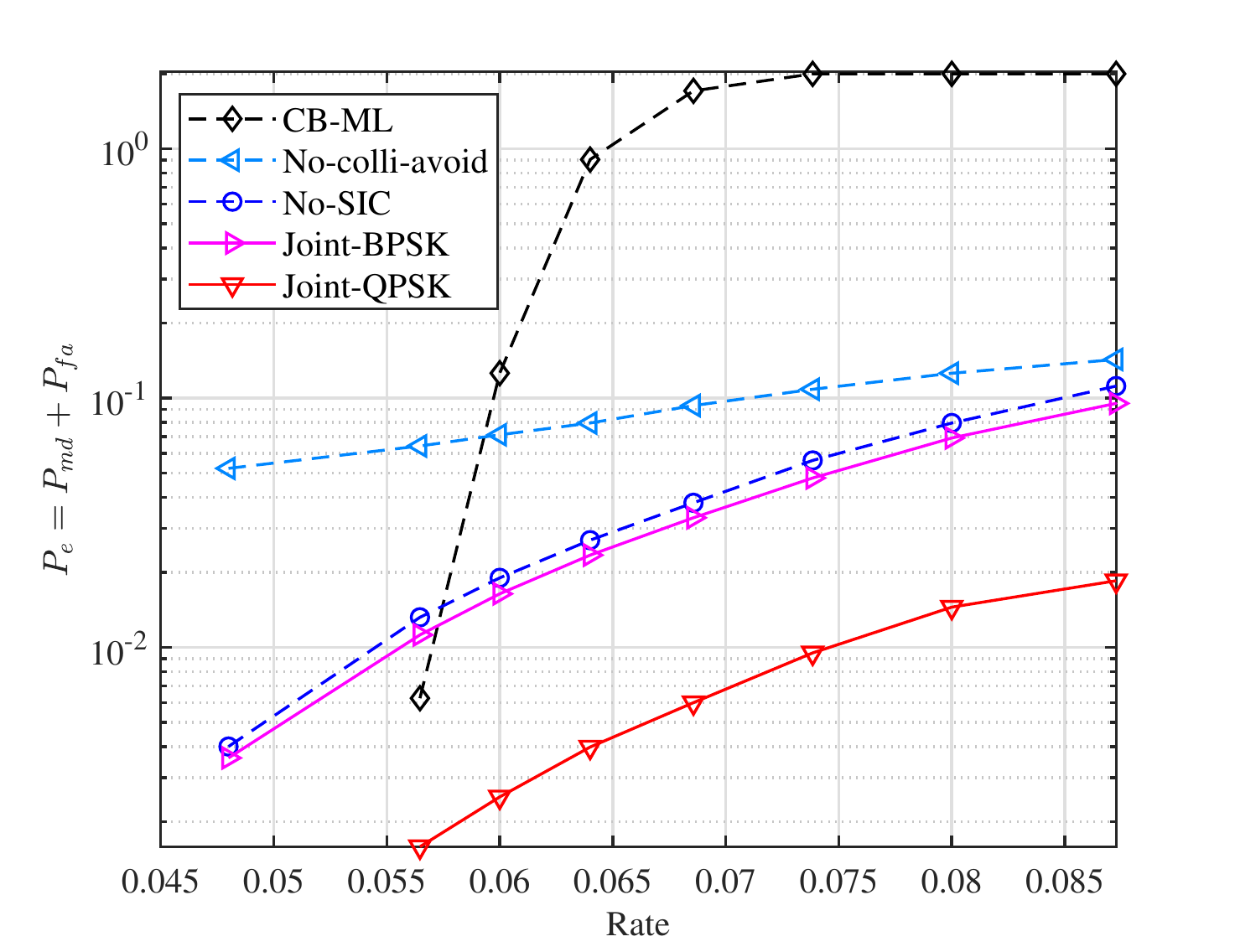}}
  	\caption{Performance of the proposed URA schemes as a function a the code rate. $E_b/N_0=10$ dB, $M=30$, $K_a=50$.}
  	\label{pic-9}
  \end{figure}

    \begin{figure*}[htpb]
	\centerline{\includegraphics[width=1.0\textwidth]{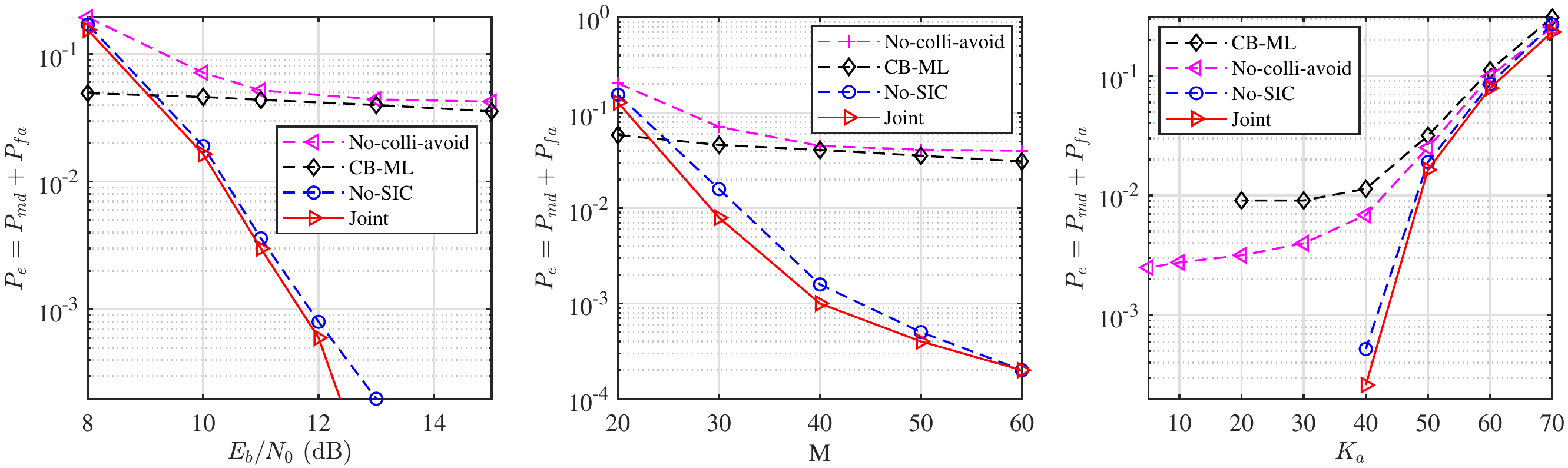}}
	\caption{Performance of the proposed URA schemes as a function of $E_b/N_0$, the number of antennas $M$ and active devices $K_a$, respectively. Parameter settings: $E_b/N_0=10$ dB, $M=30$, $K_a=50$, $L=1600$.}
	\label{pic-12}
\end{figure*}
  
 \par We first evaluate the error rate performance of the proposed schemes compared with the CB-ML scheme as a function of the code rate. The relationship between the code rate $R_c$ and channel use $L$ is $R_c=B/L$. In Fig. \ref{pic-9}, we fix the energy-per-bit $E_b/N_0=10$ dB, the number of antennas and active devices $M=30$ and $K_a=50$, respectively. In our schemes, No-colli-avoid refers to the scheme that there is no collision avoidance or the joint update. That is, the CS and LDPC phases work sequentially and potential collision may exist in the CS phase. For the other three schemes, the collision resolution protocol has been implemented and No-SIC denotes that the SIC method is not utilized in the LDPC phase nor do the two phases work jointly. Joint-BPSK and Joint-QPSK refer to the joint update algorithm with the SIC method in BPSK and QPSK modulations, respectively. As illustrated in Fig. \ref{pic-9}, there is a substantial performance enhancement compared to the CB-ML algorithm. The main reason is that the employed LDPC code has a higher code rate than the tree code proposed in \cite{Amalladinne2020tit}. As such, the proposed algorithm can work well in a relatively high rate region while the CB-ML algorithm can not. Once we set a high rate, there will be a substantial performance enhancement. However, this improvement decreases with the decrease of rate. For the target error rate $P_e=0.1$, the required code rate of the proposed Joint-BPSK is increased by $1.45$ times that of CB-ML, while the Joint-QPSK increases even more. For instance, the proposed Joint-QPSK and Joint-BPSK outperform CB-ML with a nearly $0.8$ dB gap and a $1.5$ dB gap at $R_c=0.06$, respectively. Additionally, we note that Joint-QPSK exhibits an overall $0.7$ dB performance gain compared with Joint-BPSK in terms of the error rate. This is because only half of the channel use is needed for the QPSK than BPSK modulation and thus more zeros can be padded into the sequence. Consequently, the interference of devices in the QPSK system is further reduced, resulting in improved performance. Altogether, the gain of the collision resolution protocol increases with the decrease of the code rate. However, there is no much gain for the work of joint update. This is because the gain mainly comes from the reduction of interference of devices, which has been reduced to a relatively low level by the zero-padding. As such, the joint update cannot provide more gains. Nevertheless, as we will see shortly, there is a certain gain by the joint update under a higher interference scenario.
 
 \par For the sake of complexity reduction, we employ BPSK modulation instead of QPSK in the subsequent simulations since both of them have already outperformed CB-ML. In Fig. \ref{pic-12}, we compare the error rate performance of the proposed algorithms with respect to $E_b/N_0$, the number of antennas $M$ and active devices $K_a$, respectively. The number of channel uses is fixed to $L=1600$. Correspondingly, the data is split into $16$ slots in the CB-ML algorithm, of which the parity check allocation is given in Table. \ref{tab-1}. The other parameters are set as $E_b/N_0=10$ dB, $M=30$ and $K_a=50$. As illustrated in Fig. \ref{pic-12}, the state-of-the-art method CB-ML suffers from high error floors, which stems from the poor parity check constraints. In contrast, with collision resolution, the proposed Joint and No-SIC schemes exhibit water-falling curves in terms of the error rate with respect to $E_b/N_0$ and $K_a$, while they gradually stabilize with respect to $M$. This is because the interference of devices cannot be reduced to infinitesimal by increasing $M$. As such, error still exits even for a large $M$. 
 
 \begin{figure}[htpb]
 	\centerline{\includegraphics[width=0.44\textwidth]{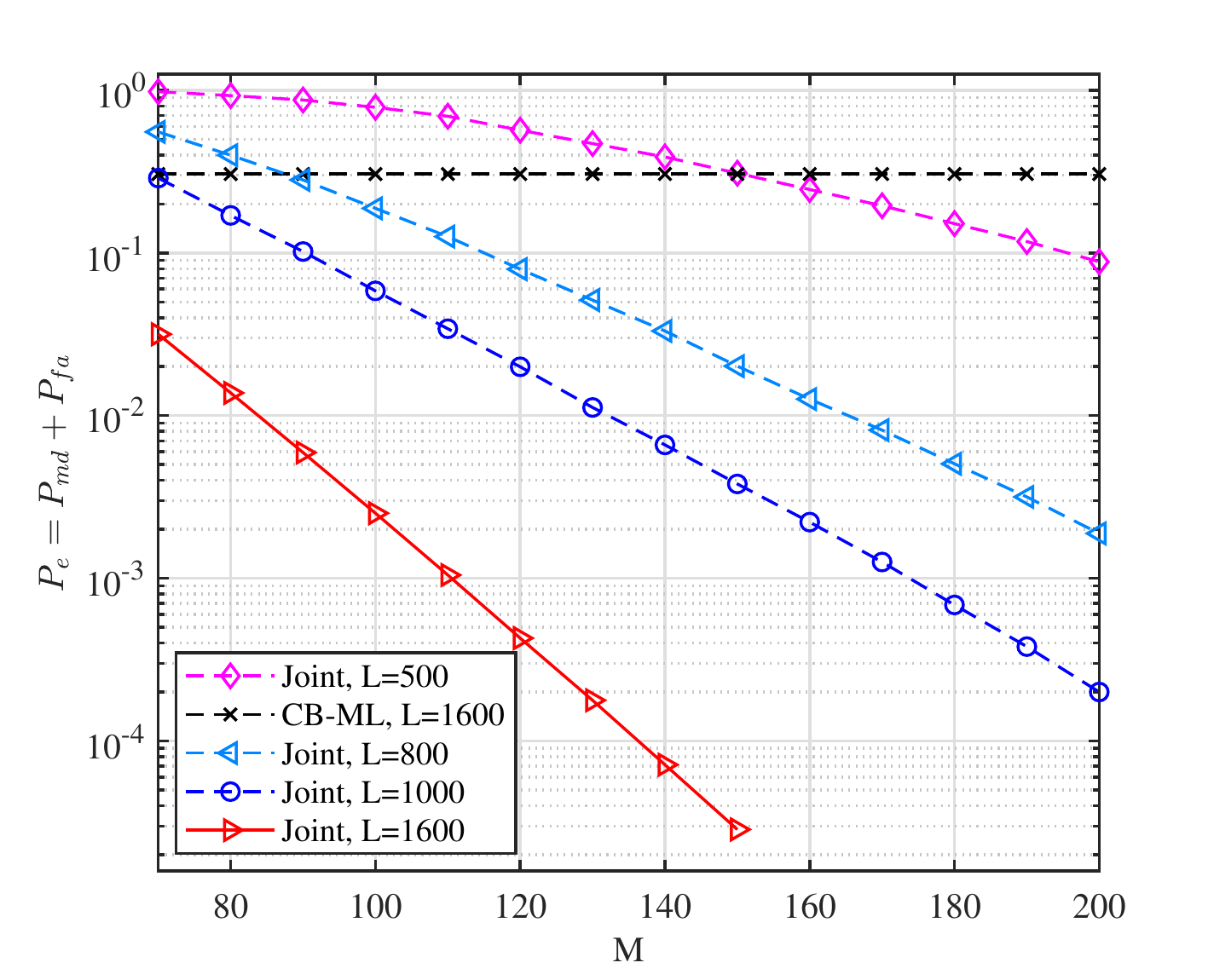}}
 	\caption{Performance of the proposed Joint-DAD-CE-DD algorithm as a function of $M$ and channel use. Parameter settings: $E_b\slash N_0 = 8$ dB, $K_a = 70$, $L_p=100$.}
 	\label{pic-13}
 \end{figure}
 
 \par Moreover, we evaluate the performance of the proposed Joint-DAD-CE-DD algorithm under a large-scale antenna with respect to different channel uses. As showcased in Fig. \ref{pic-13}, the performance of the proposed algorithm is almost linear with the antennas, but with different slopes under different channel uses. We note that CB-ML cannot work at $L=1600$, while the Joint-DAD-CE-DD algorithm can achieve $P_e < 10^{-4}$ when $M>140$. Besides, the proposed algorithm only needs half of the channel uses, i.e., $L=800$ to outperform CB-ML when $M\geq90$. To sum up, the proposed algorithms outperform CB-ML in terms of error rate and spectral efficiency, with an explicit gain with respect to various variables.
 
  \begin{figure}[h]
 	\centerline{\includegraphics[width=0.43\textwidth]{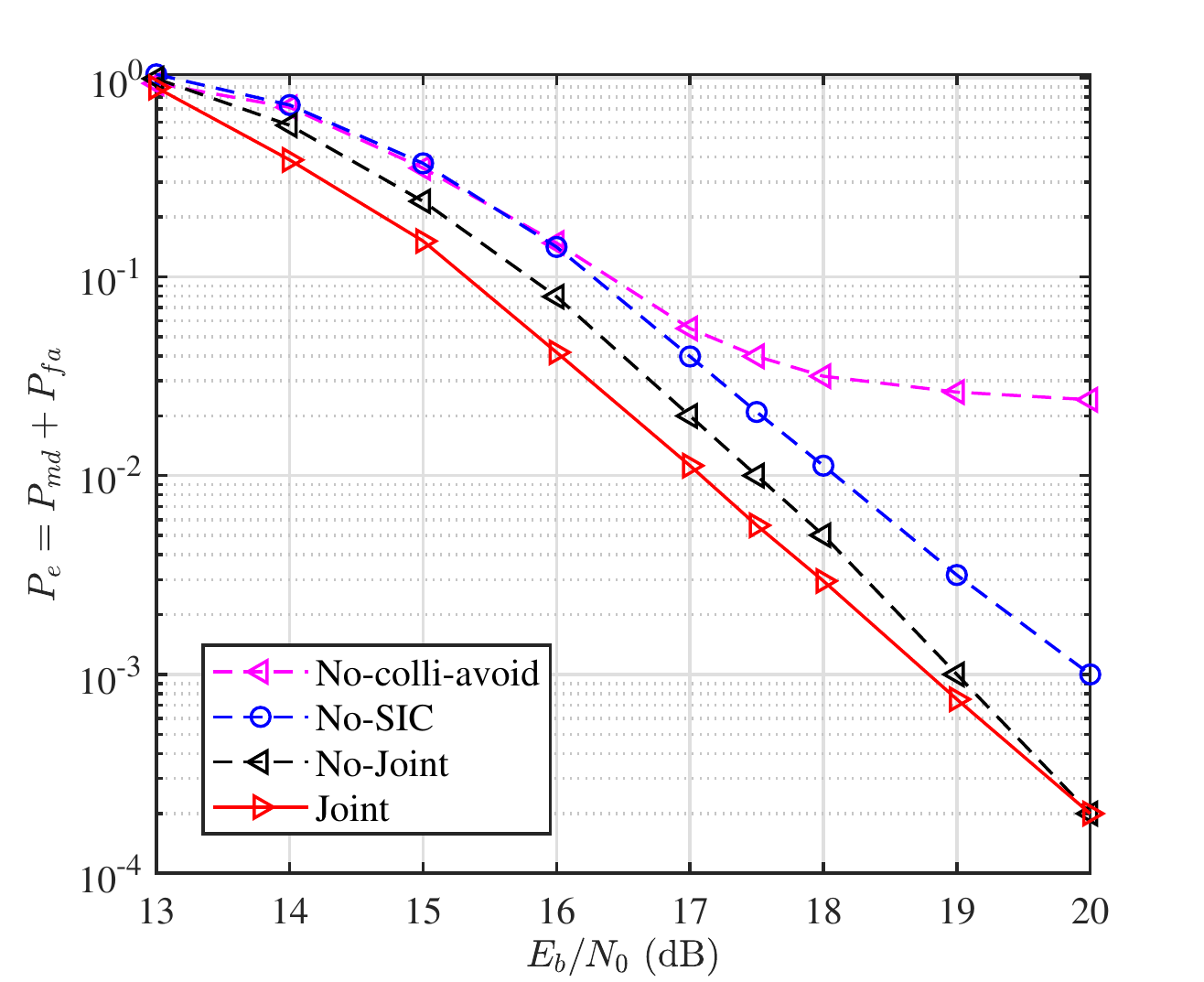}}
 	\caption{Error rate of the proposed URA schemes as a function of $E_b/N_0$. $M=30$, $K_a=40$, $L=268$, $R_c=0.36$.}
 	\label{pic-10}
 \end{figure}
 
 \par In order to provide more insights about the performance gain of different methods, we compare the error rate and CE performance among the methods in terms of $E_b/N_0$. In this regard, we fix the channel use to $L=268$ with BPSK modulation and no zero is padded, resulting in the increase of interference among devices. Besides, the code rate $R_c=0.36$, which is relatively high and the CB-ML cannot work under this circumstance. In Fig. \ref{pic-10}, No-Joint refers to the scheme with the SIC method but no joint update. Firstly, it is obvious that the collision resolution-based schemes all outperform the No-colli-avoid scheme with the increase of $E_b/N_0$. Then, the performance gain of the SIC method in the LDPC phase is about $0.3$ dB or even more when $E_b/N_0$ increases according to the comparison of No-SIC and No-Joint. Finally, by the comparison of No-Joint and Joint, there is also about $0.3$ dB gain coming from the work of joint update and it decreases with the increase of $E_b/N_0$. At a higher $E_b/N_0$, such as $20$ dB, there is nearly no performance gain by the joint update. This demonstrates that under a higher interference level of devices, the joint update algorithm can provide a certain gain, which gradually decreases with the increase of $E_b/N_0$. 
 
 \begin{figure}[htpb]
 	\centerline{\includegraphics[width=0.44\textwidth]{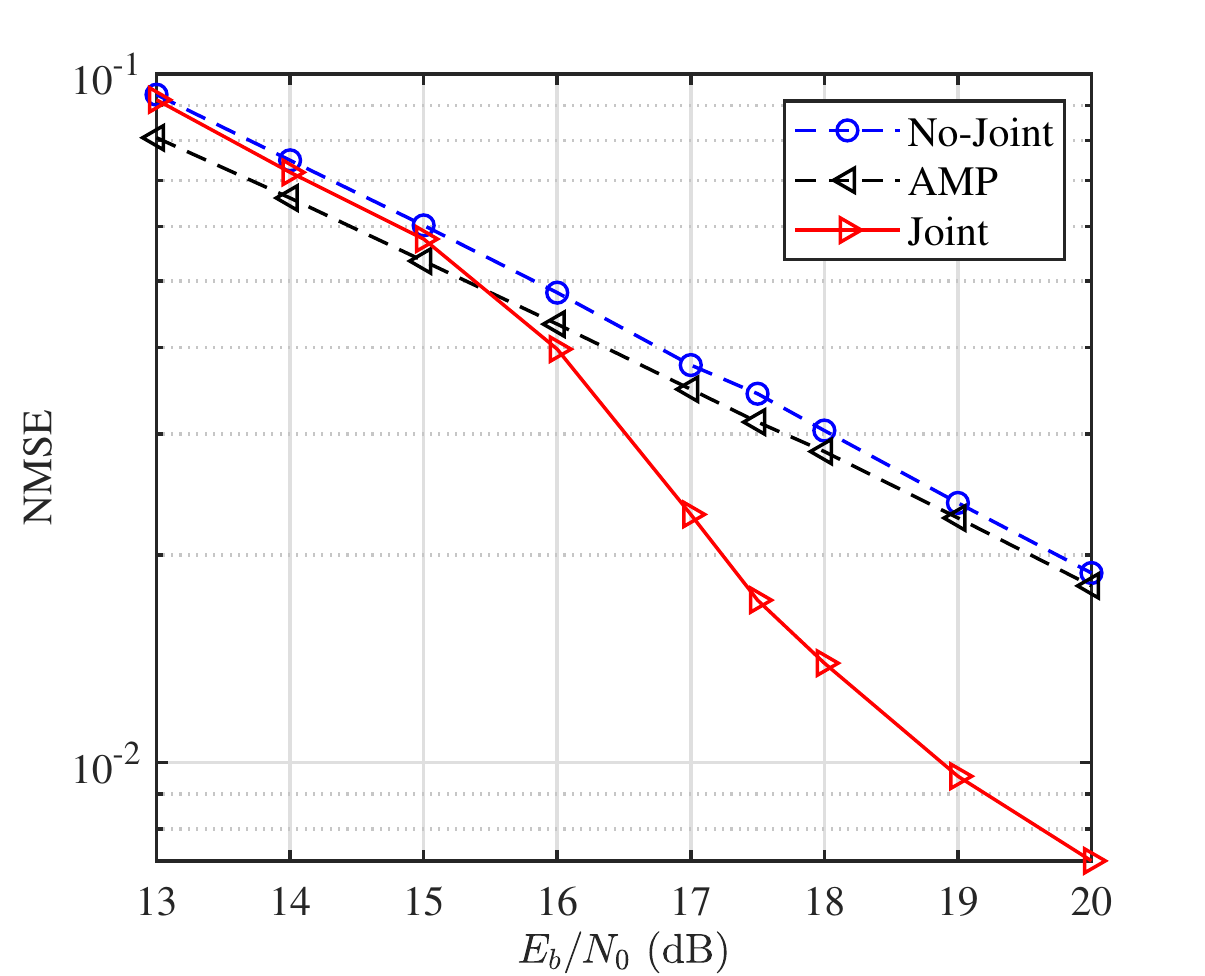}}
 	\caption{NMSE of CE with the proposed URA schemes as a function of $E_b/N_0$. $M=30$, $K_a=40$, $L=268$, $R_c=0.36$.}
 	\label{pic-11}
 \end{figure}
 
  \par The NMSE is employed to evaluate the CE performance. In Fig. \ref{pic-11}, the AMP algorithm investigated in \cite{Liu2018tsp} is utilized as a baseline for comparison. When $E_b/N_0$ ranges from $13$ to $15$ dB, the performance of Joint and No-Joint is of slight difference and both are slightly worse than AMP. However, the Joint scheme outperforms AMP with an ultra-linear speed when $E_b/N_0$ exceeds $15$ dB. As aforementioned, this gain exactly comes from the Joint-DAD-CE-DD algorithm, where the real pilot sequences as well as correctly decoded messages jointly conduct the task of CE. Figs. \ref{pic-10} and \ref{pic-11} demonstrate that the joint update indeed contributes to the improved accuracy of both the DD and CE. In general, the proposed algorithms provide substantial performance enhancements compared with CB-ML in terms of efficiency and accuracy.
 
 \section{Conclusion} \label{sec-7}
 \par In this paper, we have investigated a joint DAD, CE, and DD algorithm for MIMO massive URA. Different from the state-of-the-art slotted transmission scheme, the data in the proposed framework has been split into only two parts. A portion of the data is coded by CS and the rest is LDPC coded. Based on the principle of the BP, the iterative MP algorithm has been utilized to decode these two parts of data. Moreover, by exploiting the concept of the belief within each part, a joint decoding framework has been proposed to further improve the performance. Additionally, based on the ED and SWP, a collision resolution protocol has been developed to resolve the codeword collision issue in the URA system. In addition to the complexity reduction, the proposed algorithm has exhibited a substantial performance enhancement compared to the state-of-the-art in terms of efficiency and accuracy. The possible avenues for future work are various. More realistic channel models can be taken into consideration for MIMO URA. By utilizing the sparsity in the virtual angular domain of the spatially correlated channel, the multi-user interference can be further reduced. In addition, the proposed algorithm can be extended to handle more practical scenarios in the presence of asynchronization and frequency offset. Moreover, although this paper only studies LDPC codes, other codes, such as Turbo codes and Polar codes, could be applied if iterative soft decodings with superior performance exist in a desired  block length.
 
 \appendix
 \subsection{Analysis of the Collision Resolution Protocol} \label{append-1}
 \par Let $\bm{\phi}=\text{diag}\{\bm{\Phi}\}$ denote the vector composed of diagonal elements of $\bm{\Phi}$ and $M_p=2^{B_p}$. $\bm{\Phi}$ is the modified selection matrix defined in (\ref{equ-6-add}). Then we have $\bm{\phi} = \left[\phi_1, \phi_2, \cdots, \phi_{M_p} \right] $, which is drawn independently from the signal space
 \begin{equation}
 	\mathcal{\bm{S}}= \{ \bm{\phi}\in \{0, \cdots, K_a\}^{M_p} | \sum\limits_{i=1}^{M_p}{\phi_i}=K_a\}
 \end{equation}
according to the multinomial probability mass function :
\begin{equation}
	p(\bm{\phi}) = P\{\bm{\phi}=\left( \phi_1, \phi_2,  \!\cdots,  \!\phi_{M_p}\right)\} = \frac{K_a !}{\phi_1 ! \cdots \phi_{M_p} !}  \!\cdot \! \frac{1}{{M_p}^{K_a}} 
\end{equation}
for $\bm{\phi} \in \mathcal{\bm{S}}$. Then the probability of no collision among devices equals the probability that $K_a$ entries in $\bm{\phi}$ are one and the others are zero, which is given by
\begin{equation}
	\begin{aligned}
			P_{no-colli} &= \frac{K_a !}{{M_p}^{K_a}} \cdot \binom{M_p}{K_a} \\
			&=\frac{M_p (M_p-1)\cdots (M_p-K_a+1)}{{M_p}^{K_a}}.
	\end{aligned}
\end{equation}

\subsubsection{The first window sliding}
\par The number of devices in collision is 
\begin{equation}
	k_0 = K_a \cdot (1-P_{no-colli}).
\end{equation}
Let $n_0$ and $k_0^i$ denote the numbers of collided messages in the first $B_p$ bits and collided devices with each message, respectively, which satisfies $\sum\nolimits_{i=1}^{n_0}{k_0^i}=k_0$.
\begin{figure}[h]
	\centering
	\includegraphics[width=0.8\linewidth]{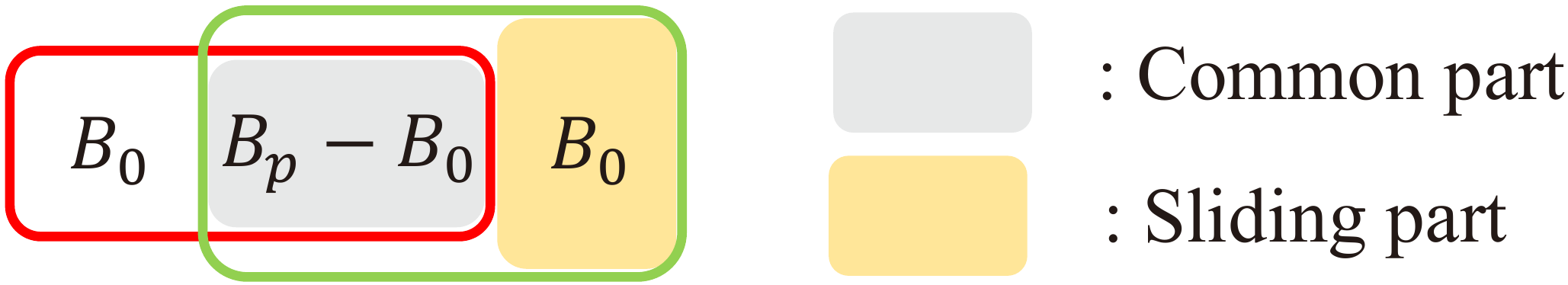}\\
	\caption{The first window sliding. Collided devices slide the window forward with window length and sliding length $B_p$ and $B_0$, respectively.}
	\label{pic-appd-1}
\end{figure}
\par  For the $i$-th collided message, the probability of no collision in the sliding part is 
\begin{equation}
	P_{no-colli-1} = \frac{M_{p1}(M_{p1} \!- \!1) \!\cdots \! (M_{p1} \!- \!k_0^i+1)}{{M_{p1}}^{k_0^i}}, i \in \left[1:n_0  \right],
\end{equation}
where $M_{p1}=2^{B_0}$. Besides, the common part is for devices to splice back the messages among different windows. If the common parts of different collided messages are the same, the message of one device will be spliced to the other's. As such, an error occurs. Therefore, this probability should also be taken into consideration. The probability that the common parts of different messages are different is
\begin{equation}
	P_{no-colli-1'} = \frac{M_{p1'}(M_{p1'}-1)\cdots (M_{p1'}-n_0^i+1)}{{M_{p1'}}^{n_0}},
\end{equation}
where $M_{p1'}=2^{B_p-B_0}$. $P_{no-colli-1}$ refers to the probability of no collision after sliding once, while $P_{no-colli-1'}$ represents the probability that messages of one device cross windows can be spliced back successfully after sliding and retransmission. However, there may still be collision after sliding once, which requires another round.

\subsubsection{The second window sliding}
\par For the $i$-th collided message, the number of devices in collision after sliding once is 
\begin{equation}
	k_1^i = k_0^i\cdot (1-P_{no-colli-1}), ~i \in \left[1:n_0  \right]. \label{equ-appd-1}
\end{equation}
Without loss of generality, we consider the $i$-th collided message in the first sliding. Likewise, let $n_1^i$ and $k_2^j$ denote the numbers of collided messages and the corresponding devices after the first sliding, respectively, which satisfies $\sum\nolimits_{j=1}^{n_1^i}{k_2^j}=k_1^i$.

\begin{figure}[h]
	\centering
	\includegraphics[width=0.8\linewidth]{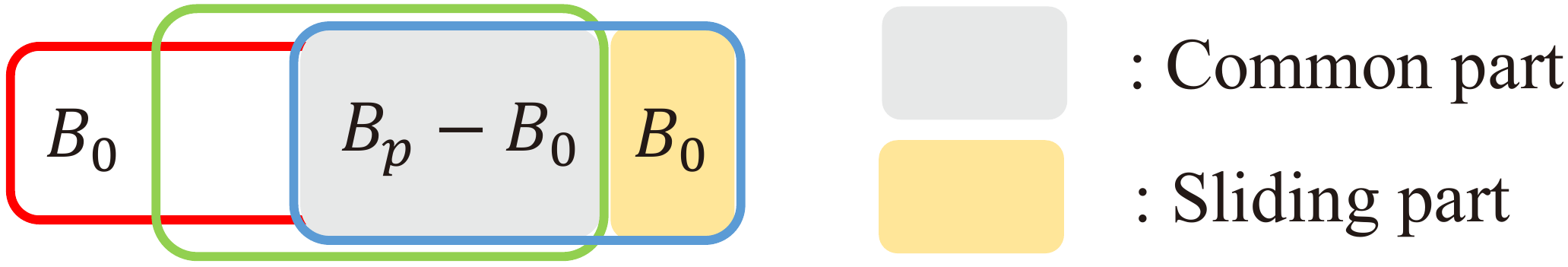}\\
	\caption{The second window sliding with window length and sliding length $B_p$ and $B_0$, respectively.}
	\label{pic-appd-2}
\end{figure}

\par For the sliding part, the probability of no collision is 
\begin{equation}
	P_{no-colli-2} = \frac{M_{p1}(M_{p1} \!- \!1) \!\cdots \! (M_{p1} \!- \! k_2^j \!+ \!1)}{{M_{p1}}^{k_2^j}}, j \in \left[1:n_1^i  \right].
\end{equation}
Note that in the common part, if there are identical sequences among $n_i^i$ messages, there will not be an error. What needs to be considered is whether the sequences among $n_0$ messages are the same or not. Thus, the probability of no collision in the common part is 
\begin{equation}
	P_{no-colli-2'} = \frac{\mathbb{P}_{M_{p1'}}^{\sum{n_1^i}}}{\prod_{i=1}^{n_0}{\mathbb{P}_{M_{p1'}}^{n_1^i}}}.
\end{equation}
Here we denote $\mathbb{P}_n^r$ as the permutations, i.e., $n!/(n-r)!$. It is obvious that $\mathbb{P}_{M_{p1'}}^{n_1^i}<({M_{p1'}})^{n_0}$ since $n_i^i<n_0$. However, the relationship between $\sum\nolimits_{i=1}^{n_0}{n_1^i}$ and $n_0$ is unknown. As such, the relationship between $P_{no-colli-2'}$ and $P_{no-colli-1'} $ is also unknown. Fortunately, since $k_2^j < k_1^i < k_0^i$, we have $P_{no-colli-2}> P_{no-colli-1}$, which demonstrates that as the window sliding progresses, the probability of devices in a collision will decrease. 

\subsubsection{The $l$-th window sliding}
\par In order to draw a general conclusion, we extend the above derivation to the $l$-th ($l>2$) window sliding. Similarly, we denote $n_{l-1}^i$ and $k_l^j$ as the numbers collided messages and corresponding devices after the $(l-1)$-th sliding, respectively, which satisfies $\sum\nolimits_{j=1}^{n_{l-1}^i}{k_l^j}=k_{l-1}^i$. As such, the probability of no collision after the $l$-th sliding is given by
\begin{equation}
		P_{no-colli-l} = \frac{M_{p1}(M_{p1} \!- \!1) \!\cdots \! (M_{p1} \!- \! k_l^j \!+ \!1)}{{M_{p1}}^{k_l^j}},
\end{equation}
where $j \in \left[1:n_{l-1}^i  \right].$ And the number of collided devices after the the $l$-th sliding is
\begin{equation}
	k_{l+1}^j = k_{l}^j\cdot (1-P_{no-colli-l}), ~j \in \left[1:n_{l-1}^i  \right]. 
\end{equation}
Obviously, we have
\begin{equation}
	k_{l+1}^j<k_{l}^j<\cdots<k_2^j<k_1^i<k_0^i<\cdots<K_a.
\end{equation}
Consequently, the probability of no collision satisfies
\begin{equation}
	P_{no-colli-l}>\cdots>P_{no-colli-1}> P_{no-colli}.
\end{equation}
In this regard, we further have
\begin{equation}
	\begin{aligned}
		k_{l+1}^j &< K_a\cdot (1-P_{no-colli-l})\cdots(1-P_{no-colli})\\
		&<K_a\cdot (1-P_{no-colli})^{(l+1)},
	\end{aligned}
\end{equation}
which demonstrates that as the sliding times $l$ goes to infinity, the number of collided devices will eventually approach zero.

\par Although the collision probability of the common part does not show an obvious trend of decreasing, in practice, we can reduce the collision probability by controlling the length of the window. There is a compromise between the collision probability of common and sliding parts. A longer sliding length will result in a lower collision probability in the sliding part, but at the same time, it will also increase that of the common part. Since the collision probability of the sliding part will eventually approach zero, the sliding length can be shorter in practice to minimize the collision probability of the common part.

\subsection{Derivation of the Gaussian and Bernoulli Messages in \ref{sec5-1}}\label{append-2}
\subsubsection{Derivation of $\bm{\Sigma}_{lk}$ at SNs} \label{append-2-1}
\par According to the formulation in (\ref{equ-15}), since the activity and channel among different devices are both independent from each other, the $(m,n)$-th $(m\neq n)$ entry for $\bm{\Sigma_{z_{lk}}}$ is given by
	\begin{equation}
	\begin{aligned} 
		&(\Sigma_{z_{lk}})_{(m,n)} \\ 
		&=\sum\nolimits_{i \in K\backslash k} {{{\left| {{A_{li}}} \right|}^2}} \mathbb{E} \left\{ {\left[ {\phi_i}{h_{im}} -\hat{\phi}_i \cdot \hat{h}_{im}\right] }\right. \\
		& \quad \quad\quad \quad\quad \quad\quad \quad ~ \left.{\cdot \left[ {{\phi_i}h_{in} } {- \hat{\phi}_i \cdot \hat{h}_{in}} \right]}^* \right\} \\  
		&=\sum\nolimits_{i \in K\backslash k} {{{\left| {{A_{li}}} \right|}^2}} \mathbb{E} \Big\{ {\left[ {{\phi_i}{h_{im}} - p_{i \to l}^{VN} \cdot \mu _{im \to lm}^{VN}} \right] } \\
		 & \quad \quad\quad \quad\quad \quad\quad \quad ~{\cdot \left[ {{\phi_i}h_{in} } {- p_{i \to l}^{VN} \cdot {{ {\mu _{in \to ln }^{VN}}} }} \right]}^* \Big\}\\  
		&= \sum\nolimits_{i \in K\backslash k}{{\left| {{A_{li}}} \right|}^2\cdot\left\lbrace \mathbb{E}\left[ \phi_i^2 \cdot h_{im}\cdot h_{in}^*\right]\right.}\\ 
		&\quad \quad\quad \quad\quad\quad  {\left. - (p_{i \to l}^{VN})^2 \cdot \mu_{im \to lm}^{VN}\cdot(\mu_{in \to ln }^{VN})^*\right\rbrace }, \label{equ-appd-2}
	\end{aligned}
\end{equation}
where
\begin{equation}
	\begin{aligned}
			&\mathbb{E}\left[ \phi_i^2 \cdot h_{im}\cdot h_{in}^*\right]= \mathbb{E} \left[\phi_i^2 \right]   \cdot  \mathbb{E} \left[h_{im}\cdot h_{in}^* \right].  \\
		& \quad = p_{i \to l}^{VN} \cdot \left[ (\Sigma_{i \to l}^{VN})_{(m,n)}+\mu_{im \to lm}^{VN}\cdot(\mu_{in \to ln }^{VN})^*\right].
	\end{aligned}
\end{equation}
Then we can rewrite (\ref{equ-appd-2}) as
	\begin{equation}
	\begin{aligned}
		&(\Sigma_{z_{lk}})_{(m,n)}\\
		& = \sum\nolimits_{i \in K\backslash k}{{\left| {{A_{li}}} \right|}^2\cdot p_{i \to l}^{VN}\cdot\left\lbrace   (\Sigma_{i \to l}^{VN})_{(m,n)} \right. } \\
		& \quad \quad \quad \quad\quad \quad~~ {\left. +(1-p_{i \to l}^{VN}) \cdot \mu_{im \to lm}^{VN}\cdot(\mu_{in \to ln }^{VN})^*  \right\rbrace} \\ 
		&= \sum\nolimits_{i \in K\backslash k}{{\left| {{A_{li}}} \right|}^2\cdot p_{i \to l}^{VN}\cdot\left\lbrace (\Sigma_{i \to l}^{VN})_{(m,n)}  \right. } \\
		& \quad \quad \quad \quad\quad \quad~~ {\left.+ q_{i \to l}^{VN} \cdot \mu_{im \to lm}^{VN}\cdot(\mu_{in \to ln }^{VN})^* \right\rbrace},  ~ m \neq n,
	\end{aligned}	
\end{equation}
where $q_{i \to l}^{VN} =  1 - p_{i \to l}^{VN}$ denotes the probability that the Bernoulli variable $\phi_k$ equals zero. Therefore, the covariance of $\mathbf{z}_{lk}$ is not diagonal. If $m=n$, we have 
\begin{equation}
	\begin{aligned}
		(\Sigma_{z_{lk}})_{(m,m)} &= \sum\nolimits_{i \in K\backslash k}{{\left| {{A_{li}}} \right|}^2\cdot p_{i \to l}^{VN}\cdot\Big\{ (\Sigma_{i \to l}^{VN})_{(m,m)}}\\ 
		&{\quad \quad \quad +~q_{i \to l}^{VN} \cdot \left| \mu_{im \to lm}^{VN}\right| ^2 \Big\}} + \sigma_n^2.
	\end{aligned}
\end{equation}
This completes the derivation.

\subsubsection{Derivation of Gaussian Messages at VNs}\label{append-2-2}
\par We give the derivations of (\ref{equ-26}) and (\ref{equ-27}) here. Rewrite the multivariate complex Gaussian pdf in (\ref{equ-23}) to canonical notation as
\begin{equation}
	f(\bm{x} \vert \bm{\mu}, \bm{\Sigma}) = \exp \left[ - \bm{x}^H \bm{\Gamma}\bm{x} + \bm{\eta}^H\bm{x}+ \bm{x}^H\bm{\eta} + \bm{\zeta} \right] 
	,
\end{equation}
where
\begin{equation}
	\begin{aligned}
		\bm{\Gamma}&= \bm{\Sigma}^{-1}, \quad \bm{\eta}=\bm{\Sigma}^{-1}\bm{\mu}\\
		\bm{\zeta} &= -\bm{\eta}^H\bm{\Gamma}^{-1}\bm{\eta} + \ln |\bm{\Gamma}|- M\ln \pi, \label{equ-appd-5}
	\end{aligned}
\end{equation}
Therefore, the product of $n$ Gaussian pdfs  is 
\begin{equation}
	\begin{aligned}
		\prod_{i=1}^n{f_i(\bm{x})}& = \exp\Big[ -\bm{x}^H\left( \sum\nolimits_{i=1}^{n}{\bm{\Gamma}_i}\right) \bm{x}  + \left( \sum\nolimits_{i=1}^n{\bm{\eta}_i}\right) ^H\bm{x}  \\
		& \quad \quad \quad  + \bm{x}^H\sum\nolimits_{i=1}^n{\bm{\eta}_i} +\sum\nolimits_{i=1}^n{\bm{\zeta}_i}   \Big], \label{equ-appd-3}
	\end{aligned}
\end{equation}
where 
\begin{equation}
	\sum\nolimits_{i=1}^n{\bm{\zeta}_i} = -\sum\nolimits_{i=1}^n{\left( \bm{\eta}_i^H\bm{\Gamma}_i^{-1}\bm{\eta}_i + ln|\bm{\Gamma}_i|\right) } - nM \ln\pi.
\end{equation}
We make a simple substitution as below
\begin{equation}
	\begin{aligned}
		\bm{\Gamma}_n &= \sum\nolimits_{i=1}^{n}{\bm{\Gamma}_i}, \quad \bm{\eta}_n = \sum\nolimits_{i=1}^n{\bm{\eta}_i}\\
		\bm{\zeta}_n  &=  -\bm{\eta}_n^H\bm{\Gamma}_n^{-1}\bm{\eta}_n + \ln |\bm{\Gamma}_n|- M\ln \pi.
	\end{aligned}
\end{equation}
Then we can rewrite (\ref{equ-appd-3}) as
\begin{equation}
	\begin{aligned}
		\prod_{i=1}^n{f_i(\bm{x})}& = \exp \Big[ -\bm{x}^H \bm{\Gamma}_n\bm{x}  + \bm{\eta}_n^H\bm{x} + \bm{x}^H\bm{\eta}_n +\bm{\zeta}_n  \\
		& \quad \quad \quad +\sum\nolimits_{i=1}^n{\bm{\zeta}_i} -\bm{\zeta}_n \Big] \\
		&= c \cdot \exp\left[ -\bm{x}^H \bm{\Gamma}_n\bm{x}  + \bm{\eta}_n^H\bm{x} + \bm{x}^H\bm{\eta}_n +\bm{\zeta}_n\right], \label{equ-appd-4}
	\end{aligned}
\end{equation}
where $c=\sum\nolimits_{i=1}^n{\bm{\zeta}_i} -\bm{\zeta}_n$ is a constant for normalization. Since the $n$ Gaussian pdfs are independent, the product of which still follows the Gaussian distribution. Comparing with (\ref{equ-appd-5}) and  (\ref{equ-appd-3}), the covariance and mean of the Gaussian distribution in (\ref{equ-appd-4}) are
\begin{equation}
	\begin{aligned}
		\bm{\Sigma}_n &= \bm{\Gamma}_n^{-1} = \left( \sum\nolimits_{i=1}^{n}{\bm{\Sigma}_i^{-1}} \right) ^{-1}\\
		\bm{\mu}_n &= \bm{\Sigma}_n \cdot  \sum\nolimits_{i=1}^n{\bm{\eta}_i}= \bm{\Sigma}_n \cdot  \sum\nolimits_{i=1}^n{\bm{\Sigma}_i^{-1}\bm{\mu}_i}. \label{equ-appd-6}
	\end{aligned}
\end{equation}
Similarly, in \eqref{equ-25}, it is the product of Gaussian pdfs $\left\lbrace f(\bm{h} \vert \bm{\mu}_{i \to k}^{SN}, \bm{\Sigma}_{i \to k}^{SN}), i \in \mathcal{L} \backslash l\right\rbrace $ and $f(\bm{h} \vert \bm{\mu}_k^{pri}, \bm{\Sigma}_k^{pri})$. According to the rule in (\ref{equ-appd-6}), we can obtain \eqref{equ-26} and \eqref{equ-27}.

\end{document}